\documentclass[fleqn,usenatbib,useAMS]{mnras}

\usepackage{scicite}  
\usepackage{times}  
\usepackage{graphicx}
\usepackage{amsmath} 
\usepackage{amssymb}
\usepackage{rotating} 
\usepackage{color}
\usepackage{float} 
\usepackage{multicol} 
\usepackage{bm} 
\usepackage{pdflscape} 

\newcommand{\rsrc}{R_{\rm X}}
\newcommand{\rfront}{f_{ec}}
\newcommand{\vel}{v_{pl}}
\newcommand{\tmid}{T_{\rm mid}}
\newcommand{\bimpact}{b}
\newcommand{\ctsrc}{c_{\rm X}}

\usepackage{newtxtext,newtxmath}

\usepackage[T1]{fontenc}

\DeclareRobustCommand{\VAN}[3]{#2}
\let\VANthebibliography\thebibliography
\def\thebibliography{\DeclareRobustCommand{\VAN}[3]{##3}\VANthebibliography}

\usepackage{graphicx}	
\usepackage{amsmath}	
\usepackage{amssymb}	

\title[M51-ULS-1b]{M51-ULS-1b: The First Candidate for a Planet in an External Galaxy}

\author[R.Di Stefano et al.]{
R. Di Stefano,$^{1}$\thanks{E-mail: rdistefano@cfa.harvard.edu}
Julia Berndtsson,$^{2}$
Ryan Urquhart,$^{3}$
Roberto Soria,$^{4, 5}$
Vinay L.\ Kashyap,$^{1}$
\newauthor Theron W. Carmichael,$^{1}$
Nia Imara$^{1}$\\
$^{1}$Center for Astrophysics $|$ Harvard \& Smithsonian, 60 Garden St., Cambridge, MA 02138, USA\\
$^{2}$Princeton University, Princeton, NJ 08544, USA.\\
$^{3}$Center for Data Intensive and Time Domain Astronomy, Department of Physics and Astronomy, Michigan State University, East Lansing MI 48824, USA\\
$^{4}$University of the Chinese Academy of Sciences,
Beijing 100049, China\\
$^{5}$Sydney Institute for Astronomy, School of Physics A28, Sydney, NSW 2006, Australia\\
    }

\date{Accepted XXX. Received YYY; in original form ZZZ}

\pubyear{2020}

\begin{document}
\label{firstpage}
\pagerange{\pageref{firstpage}--\pageref{lastpage}}
\maketitle

\begin{abstract}
Do external galaxies host planetary systems? Many lines of reasoning suggest that the answer must be ``yes''.  In the foreseeable future, however, the question cannot be answered by the methods most successful in our own Galaxy.
We report on a different approach which focuses on bright X-ray sources (XRSs).  M51-ULS-1b is the first planet candidate to be found because it produces a full, short-lived eclipse of a bright XRS.  M51-ULS-1b has a most probable radius slightly smaller than Saturn.  It orbits one of the brightest XRSs in the external galaxy M51, the Whirlpool Galaxy, located 8.6 Megaparsecs from Earth.  It is the first candidate for a planet in an external galaxy.  The binary it orbits, M51-ULS-1, is young and massive. One of the binary components is a stellar remnant, either a neutron star (NS) or black hole (BH), and the other is a massive star.   X-ray transits can now be used to discover more planets in external galaxies and also planets orbiting XRSs inside the Milky Way.
\end{abstract}








\date{}




\baselineskip12pt

\maketitle

\section{Introduction}
Planets are ubiquitous in the Milky Way. The conditions under which the known planets formed exist in other galaxies as well. Yet each external galaxy occupies such a small area of the sky that the high projected stellar density makes it difficult to study individual stars in enough detail to detect the signatures of planets through either radial velocity measurements or transit detection, the two methods responsible for the discovery of more than 4300 exoplanets  (exoplanet.eu).

External galaxies host relatively small numbers (a handful to several hundred) of bright X-ray sources (XRSs). Luminous XRSs in external galaxies can therefore be spatially resolved, and we can measure the count rate and X-ray flux from each XRS as a function of time (i.e., deriving the ``light curve'').  
The dominant set of bright XRSs in external galaxies are 
X-ray binaries (XRBs) in which black holes (BHs) or neutron stars (NSs), the remnants of massive stars, accrete matter from a stellar companion. 

\citet{2018ApJ...859...40I} suggested that XRBs may be ideal places to search for planets, because the cross-sectional areas of the X-ray emitting regions can be comparable to or even smaller than planetary cross sections. 
A planet passing in front of the X-ray emitting region may produce a total or near-total eclipse of the X-rays \citep{2018ApJ...859...40I}.
Furthermore, we know that planets are likely to inhabit XRBs. For example, studies of radio emission from four NSs that spin at millisecond periods ({\sl recycled pulsars}) and which were previously XRBs, have led to the discovery of planets \citep{WOLSZCZAN20122}.
We therefore expect active XRBs to also host planets
On a related note, eclipse timing variations
in the
XRB MXB~$1658-298$, suggest the presence  of a 23 Jupiter-mass object, a {brown dwarf}\footnote{Planets have masses below roughly $13\, M_J$, where $M_J$ is Jupiter's mass; brown dwarfs have masses between that and $\sim 0.075\, M_\odot=75\, M_J$, the lower mass limit for stars.}, in a 1.5 AU orbit \citep{Jain_Paul_etal}.

M51-ULS-1b is the first planet candidate discovered when it passed in 
front of an XRS whose size is comparable to its own. It completely blocked the X-rays from the XRB M51-ULS-1 for a time interval of 20-30 minutes, with the excursion from baseline lasting roughly 3 hours. There were no simultaneous observations in the optical or infrared, but the regions emitting at these longer wavelengths are so large in comparison with the XRS that there would likely not have been a detectable decline in flux.
This phenomenon is to be contrasted with planetary transits of {\sl stars}, which produce relatively small dips in flux across wavebands. During stellar transits the dip in X-ray flux is ${\cal O}(1\%)$, rather than ${\cal O}(100\%)$. The specific shape  of the stellar-transit X-ray dip can be modeled by the interactions of X-rays from the stellar corona with the planetary atmosphere. \citep{HD18933b_in_Xrays}.   

The method of X-ray transits we discuss here applies to XRBs rather than stars, and can produce a full eclipse.  During the transit of M51-ULS-1, a candidate planet passes in front of a soft-X-ray source that has an effective radius of $\sim 1/3\, R_J$, where $R_J$ is the radius of Jupiter.  The method can be applied to external galaxies like M51, because the field of view of the present generation of X-ray detectors is large enough to encompass dozens of bright XRSs, and the total exposure times extend to about $1$~Ms ($\sim 11.6$~d).  XRSs in the Local Group and in the Milky way, which are generally intrinsically dimmer, can be studied as well. When applied to them, the method is sensitive to planets in closer orbits in which the transits repeat on shorter time scales.

There is a special excitement to discovering planets in external galaxies. The identification of X-ray transits is the only method in which host stellar systems at distances of Mpc to tens of Mpc  can be unambiguously identified.  Microlensing is sensitive to planetary masses in external galaxies, and possible free-floating planets found through quasar microlensing have been reported \citep{2018ApJ...853L..27D}. Microlensing of light from stars in galaxies such as M31 can also lead to planet discoveries, and in the era of LSST, deep drilling in some crowded fields combined with the survey's planned image differencing analysis (so-called pixel lensing), should discover planets that orbit intervening stars
\citep{2009MNRAS.399..219I}. The identification  of the host star can, however, range from difficult to impossible. In fact, for only a small number of the 120 Milky-Way planets discovered via microlensing is information about the host star available (exoplanets.edu).  
For planets discovered via X-ray transits, we know and can study the system orbited by the star, even when only a single transit is detected.  We can, for example, identify the range of orbital separations the candidate planet has from the XRS, and study the feasibility of its survival within its present environment as well as its survival during previous stages of the binary's evolution.  

The discovery of M51-ULS-1b initiates investigations of planets and substellar masses orbiting massive stars, which have proved difficult to discover, with only a handful of the known exoplanets orbiting stars with masses larger than $2-3\, M_\odot$.
High-mass stars (those with mass greater than about $10\, M_\odot$) tend to be formed only in binaries or in systems with higher-multiplicities 
\citep{PQ}. Binary-high-mass stars can undergo interesting evolutions that lead to a range of energetic hydrogen-poor supernovae and eventually to BH-BH, NS-NS, or BH-NS mergers.
The candidate planet we have discovered is in a circumbinary orbit around a system experiencing an intermediate phase of evolution. One star has evolved and is now a BH or NS and its companion is a massive star donating matter,  making the compact object highly luminous. With a luminosity $> 10^{39}$~erg~s$^{-1}$, M51-ULS-1 is an {\sl ultraluminous} X-ray source.  Its ultimate fate depends on the mass of the donor relative to that of the accretor. The discovery of planets around such a system expands the realm of known planetary environments.

In \S2 we describe our search through archived X-ray data for X-ray transits, and the identification of a transit of the XRB M51-ULS-1. Section 3 is devoted to establishing the properties of the XRB, which are subsequently used to better understand the planet candidate and its orbit. One of the properties of the XRS that is particularly useful is that its spectrum is thermal, so that we can derive the effective radius of the XRS.  Optical observations establish that the XRB is young, likely younger than $20$~Myr.    To ensure that the transit is not simply an example of a commonly found type of X-ray dip, we compare it in \S 4 with other dip-like events. We find that the shape and spectral evolution of the transit are different from those of other flux dips. In particular the constancy of the spectrum is what is expected during an eclipse or transit rather than obscuration due to gas and dust, or to a state change.  In \S 5 we establish that the light curve is well fit by a transit model which yields the size of eclipser relative to the XRS as well as the relative speed. In \S 6 we explore the nature of the eclipser. We find that the eclipser is substellar and is most likely to be a planet. In \S 7 we study the candidate planet's orbit. We find that it is wide enough to expect that a planet could survive in the radiation field of the XRB, and also to suggest that a planet could have survived the prior evolution of the XRB. Section 8 considers the implications of the discovery, specifically how large a population of planets could inhabit the set of XRBs we studied?  and what are the prospects for future detections?

\begin{figure*}[ht!]
\begin{center}
\includegraphics[width=0.75\textwidth]{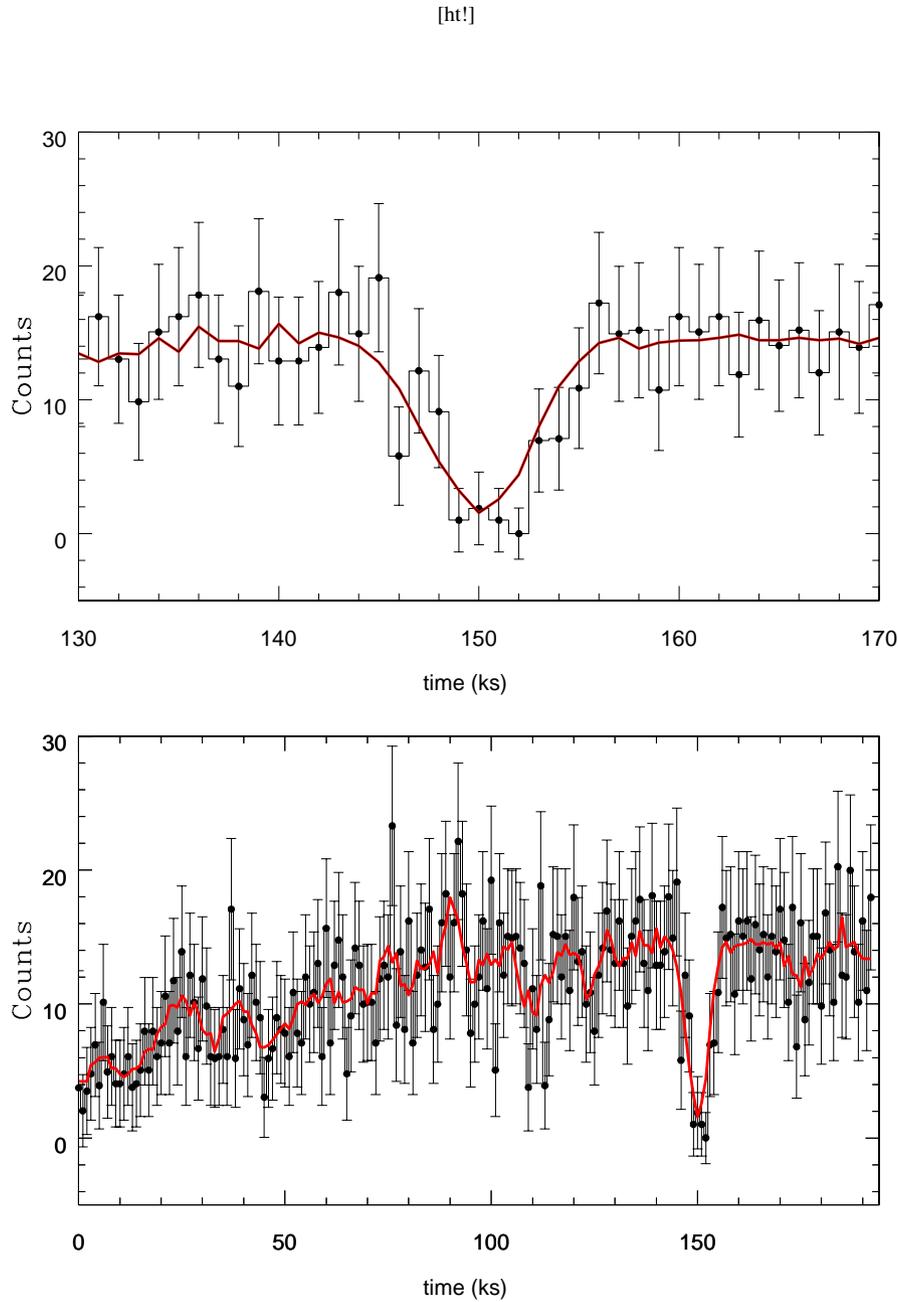} 
\caption{Background-subtracted X-ray light curves defined by data points for \textit{Chandra} ObsID 13814. {\sl Black:} counts in 1~ks bins, and the associated 1-$\sigma$ uncertainties. {\sl Red:}  running average computed over a timescale of $\pm$2~ks. 
{\sl Horizontal axis:} time in ks; 
{\sl Vertical axis:} number of counts per bin. {\bf Top panel:} the short duration eclipse and roughly 20 ks on each side. {\bf Bottom panel:}  the entire duration of the observation.}
\label{fig:1}
\end{center}
\end{figure*}

\section{Search for X-Ray Transits}

We conducted a systematic search for possible transits in the {\sl Chandra} X-ray light curves of XRSs in three galaxies: M51 (a face-on interacting late-type galaxy), M101 (a face-on late-type galaxy), and M104 (an edge-on early-type galaxy with some star formation). 
These light curves were available because they had recently been studied
for other purposes \citep{2018MNRAS.477.3623W,
2016ApJ...831...56U,
2016MNRAS.456.1859U}.
It is possible to make discoveries of planetary transits in archived X-ray light curves, because short-term time variability was often not a primary focus of the
original observing programs. Even stellar eclipses lasting ten or more hours have  been found after
the initial analyses were complete. Planetary transits, a phenomenon that has apparently not been previously targeted, exhibit short-duration deficits of photons, and are therefore particularly prone to be missed or misidentified.

We considered all observations of duration greater than $5$~ks, and all obsids (individual observations) for each XRS observed to have had a flux corresponding to 
$L_X[0.5\, {\rm keV} -8\, {\rm keV}]>10^{37}$~erg~s$^{-1}$ during at least one observation. We studied 667 light curves produced by 55 XRSs in M51, 1600 light curves from 64 XRSs in M101, and 357 light curves from 
119 XRSs in M104.  
The numbers of light curves are larger than the total numbers of XRSs because each physical source was in the fields of multiple exposures. 

We conducted an automated search specifically designed to identify transits. We required only that there be at least one 1-ks interval with no X-ray counts, and that, however long the low state lasted, there should be a baseline with roughly equal count rates prior to and after the downward dip.  We applied our search algorithm to all 2624 light curves in the sets described above.  

The criteria, that the light curve should exhibit a drop to zero measured count rate in at least one 1-ks bin, and that the downward deviation should start from and return to a baseline, were enforced as follows.  Considering an individual light curve, and denoting the counts in bin $i$ as $C(i),$ we identified all values of $i$ for which $C(i)=0.$  We then considered the time bins just before $i$
($C(i-j)$, $j=1,2,...$) and just after ($C(i+j)$, $j=1,2,...$). The purpose 
of this was to measure the duration of the interval during which the count rate was consistent with zero. We did this by counting the total number of consecutive bins in which the count rate was equal to or smaller than 1\footnote{Given the uncertainty associated with small numbers of counts, this was a strict criterion. In trials where we relaxed it, the low states were often clearly associated with longer-term variability rather than with well-defined events.}.
The first and last of these bins were, respectively, $i_{low}$ and $i_{high}$, so the duration of the low state is
$[(i_{high}-i_{low})+1]$~ks. To determine whether the low state corresponds to a transit, we needed to establish whether the dip started from and returned to a baseline. We therefore considered, in turn, four pairs of points: 
$({\cal C}_1=C(i_{low}-k), {\cal C}_2=C(i_{high}+k)),$ 
where the value of $k$ ranges from $1$ to $4$. For each of the four pairs we defined $\sigma=\sqrt{max({\cal C}_1,{\cal C}_2)}$ If the absolute value of the difference between ${\cal C}_1$ and ${\cal C}_2$ was less than $2\, \sigma$, we considered that pair to be a match. We also required that both ${\cal C}_1$ and ${\cal C}_2$ be $7$ or larger, to ensure that the count rate at baseline is significantly higher than it would have been during the low-count-rate interval.  We conducted this check for four
pairs of points ($k=1,2,3,4$).  If at least two of the four pairs had high enough count rates and were also matches, we considered the event to be a possible transit and flagged it for visual inspection. The 2624 light curves in our study yielded one interval for inspection. This was the light curve in Figure \ref{fig:1}, with an apparent transit lasting $10$~ks to $12$~ks.  

\begin{figure*}
 \begin{center}
    \includegraphics[width=0.42\linewidth]{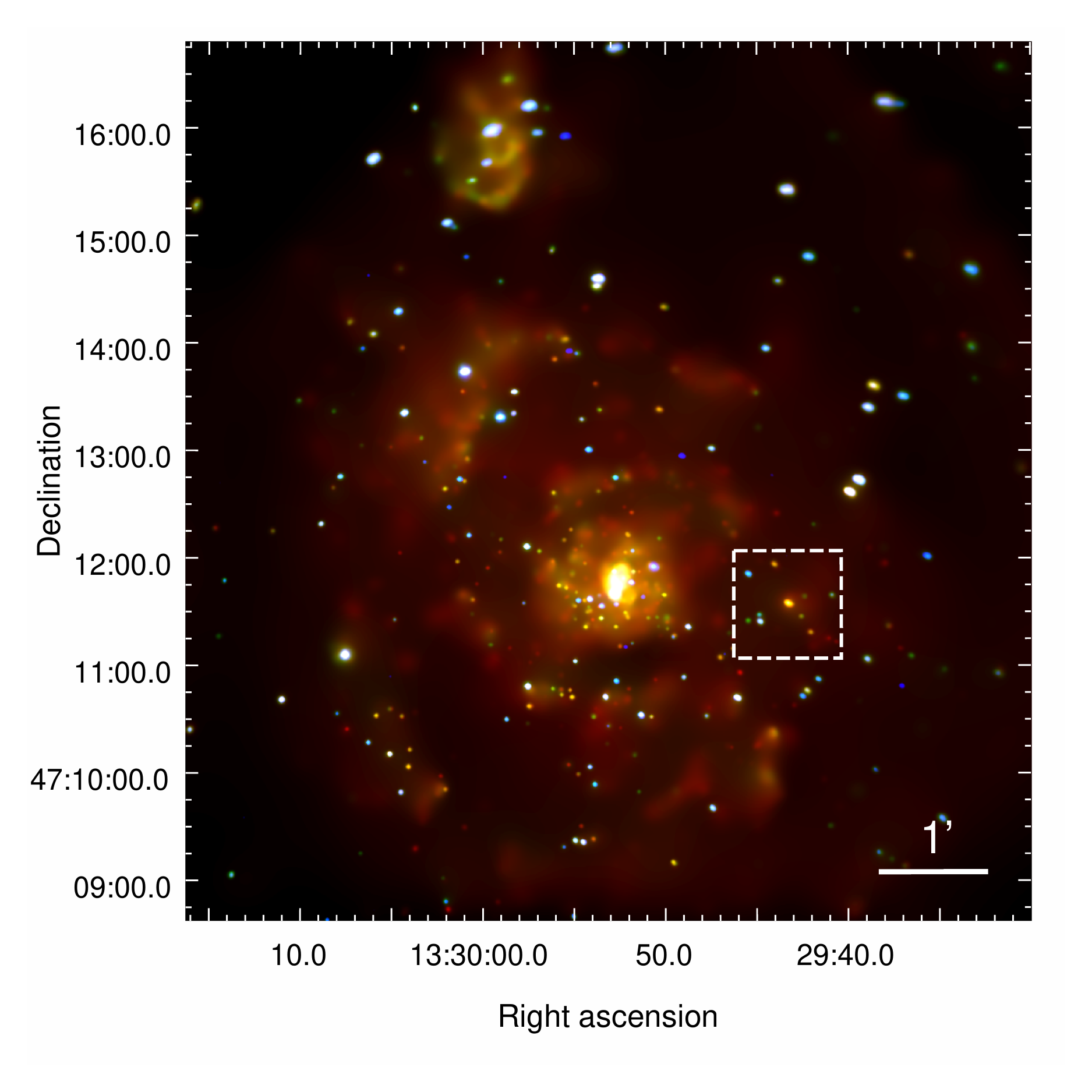}
    \includegraphics[width=0.42\linewidth]{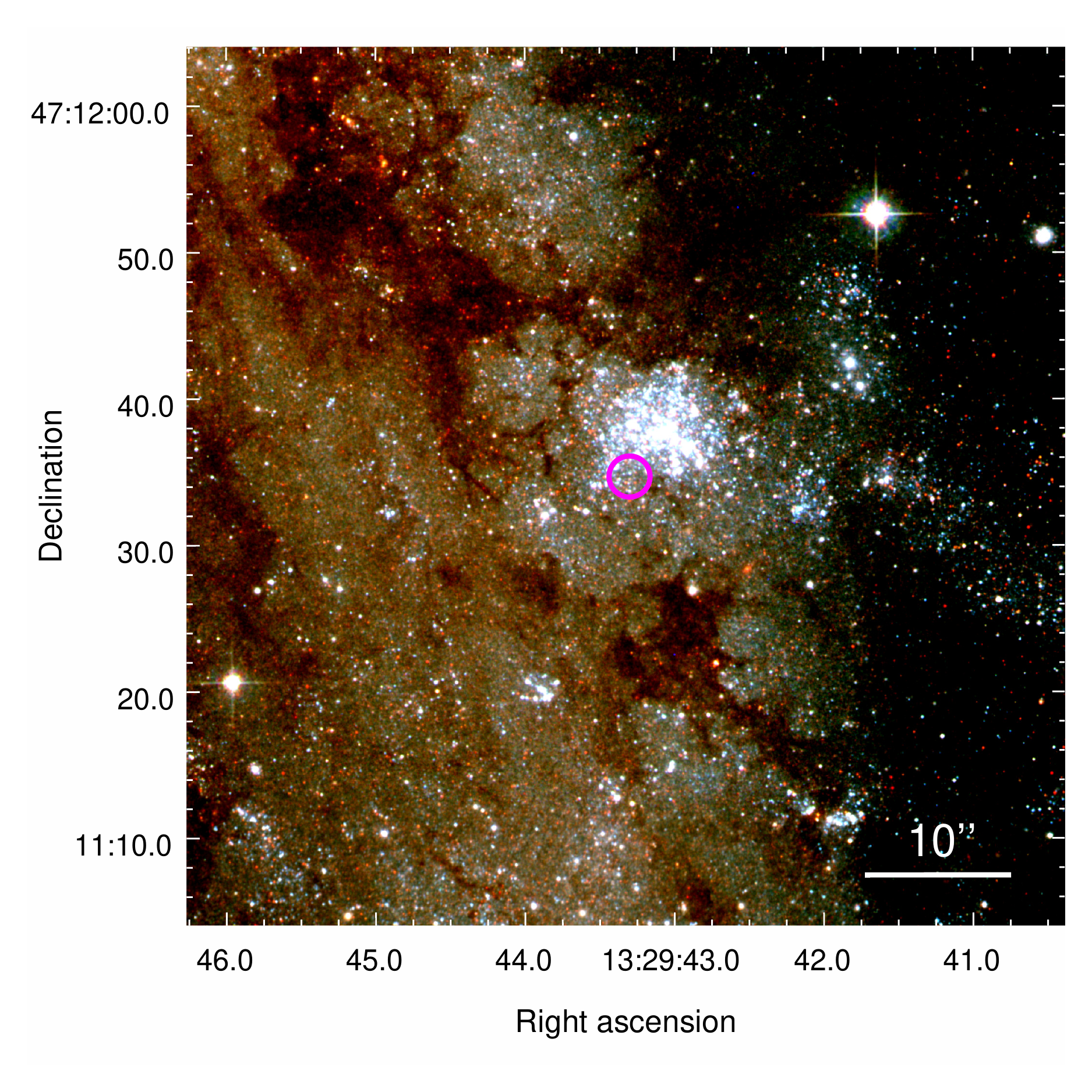}
    \caption{Left: false RGB stacked \textit{Chandra}/ACIS-S image of the Whirlpool Galaxy, M\,51 (total exposure of $\approx$850\,ks). Colored points are XRSs: Red is 0.3-1\,keV; green is 1-2\,keV; blue is 2-7\,keV. M\,51-ULS1 is the orange source at the center of the $60^{\prime\prime}\times60^{\prime\prime}$ dashed white box. Diffuse emission is from hot gas. Right: \textit{HST} image of the area defined by the white box in the top image. Red is the F814W band; green is F555W; blue is F435W. The magenta circle marks the X-ray position of M\,51-ULS, which lies at the edge of a young star cluster. The source is located at right ascension and declination 13:29:43.30, +47:11:34.7, respectively.}
    \label{fig:m51_image}
 \end{center}
\end{figure*}

We also employed other approaches to study the light curves. 
For each of the 2624 light curves, we plotted the cumulative count rate, using a method developed by \citet{2017Sci...355..817I}  to discover flares in XRSs. We tested the method to determine whether it could also discover dips, and found it to be effective. The signature of an eclipse, for example, is a flat region in the plot of cumulative count rate versus time. 

We also measured the total number of counts in each exposure, $C$, and used $C/T_{exp}$, where $T_{exp}$ is the exposure time, to compute the average count rate. We 
computed running averages of the counts per ks, located
the positions of local extrema, and binned the data, initially selecting the bin size so that there would be an average of 10 counts per bin. We subsequently conducted a visual inspection of each light curve with $C>100$ to look for dips in flux. We compared the results of our algorithmic analyses (e.g., significance of changes in flux)  with visually identifiable features in the light curve. This process led to the selection of the event shown in Figure \ref{fig:1}. In the appendix we present the details of the {\sl Chandra} and {\sl XMM-Newton} data sets we used to study M51.  Data employed for our searches in M101 and M81 was used in a similar manner by previous studies and more details can be found in \citep{2018MNRAS.477.3623W,
2016ApJ...831...56U,
2016MNRAS.456.1859U}.

That our thorough search through the {\sl Chandra} light curves of M51, M101, and M104 identified a single transit candidate, demonstrates both that transits can be found, and also that transit profiles are not common features of the light curves of extragalactic XRSs.  Criteria that are less strict (e.g., not requiring a drop to zero flux) might have identified more candidates; our goal, however, was to identify only strong candidates that could then be subjected a sequence of further tests. 

As expected for the transit of a spherical object with a well-defined edge, the dip is roughly symmetric and, as we will show in \S 4, the features of the transit are consistent for photons with different energies.  The transit has the characteristic shape expected when the size of the transiting object is similar to that of the background source. Fortunately, X-ray studies of the binary, described below, provide an estimate of the size of the XRS.

\section{The X-Ray Binary M51-ULS-1}

\subsection{X-Ray Properties}
The X-ray source exhibiting the apparent transit is  M51-ULS-1, one of the brightest XRSs in M51, located at right ascension and declination 13:29:43.30, +47:11:34.7, respectively.
The X-ray luminosity (0.3--7 keV) of M51-ULS-1 is $\sim 10^{39}$~erg~s$^{-1}$ \citep{2016MNRAS.456.1859U}, roughly $10^5 - 10^6$ times brighter in X-ray emission than is the Sun at all wavelengths combined. The high X-ray luminosity ensured that the count rate was large enough to both identify and study the transit. 

M51-ULS-1 belongs to a subclass of fast-accreting XRSs known as {\it ultraluminous supersoft sources} (ULSs), characterized by high luminosity and an almost purely thermal spectrum with typical blackbody temperature $\sim 100$~eV and emitting radius $\sim 10^9$~cm \citep{2016MNRAS.456.1859U}. 

The data exhibiting the short eclipse were collected during a 190-ks  Chandra pointing (ObsID 13814, 2012 September 20). During that observation, the average effective radius of the X-ray-emitting region was estimated to be $\rsrc = 2.5 ^{+4.1}_{-1.1} \times 10^9$~cm (90\% confidence interval) \citep{2016MNRAS.456.1859U}.  
The radius is extracted from a fit to the broadband X-ray data collected just prior to and after the dip to zero flux. 
\footnote{If the underlying spectrum is not a blackbody, the size of the X-ray emitting region could be somewhat smaller or larger in a way not accounted for in the uncertainty limits.}

In this particular case, the radius, $R_X$ of the transited XRS is on the order of a few $\times 10^9$~cm, consistent with sizes of known planets.
In \S 5 we find that the light curve data are well fit by a transit model. The model yields a range of eclipser radii in units of $R_X$, as well as a range of relative velocities.  

\subsection{Optical Properties and Age}
Optical observations of M51  provide clues to the age of M51-ULS-1,
 a good indicator of the age of M51-ULS-1b, since the latter is likely to have been formed with the binary, or else in the binary's natal cluster.  Alternatively, 
if it formed as a result of the evolution of the binary, or one of its components, M51-ULS-1b would be younger.

Several lines of evidence suggest that M51-ULS-1 is a young system. 
The most direct evidence is from \citep{Terashima} who have identified a possible counterpart in an HST image, consistent with stellar type B2-8. Comparison with the relevant isochrones yield an age range estimate of 4~million~yr to 16~million~yr.  While there is a probability of $0.17$ that a star this bright or brighter would be present by chance, we note that a bright counterpart is expected because a donor that can give mass at a rate high enough to produce an accretion luminosity of $10^{39}$~erg~s$^{-1}$ must either be massive or highly evolved, and would be bright in either case.  The fact that there is no bright red star in the vicinity
argues for a massive star, with the colors consistent with those of a blue supergiant.  It is possible, however, that the counterpart includes light from the accretion disk, altering the age estimate.

While the counterpart suggests an age smaller than about 20~million~yrs, other considerations independently constrain the age to be less than $\sim 10^8$~yrs.  For example, the  HST image shows that  M51-ULS-1 is located on the edge of a young stellar cluster surrounded by diffuse H$_\alpha$ emission (Figure 2).
 Furthermore, \citep{2017MNRAS.466.1019S} place M51-ULS-1 in a spiral arm. They and other authors have identified M51-ULS-1 a high-mass X-ray binary, indicating that the donor is young.  Finally, ULSs are preferentially found in young stellar populations.
\citep{2016MNRAS.456.1859U}.
The preferred age of the system is less than about 20~Myr, and its maximum age is roughly $10^8$~yr.

\subsection{Binary Properties}

The total mass, $M_{tot}$, of the XRB. 
is the sum of the accretor's mass, $M_a$ and the donor's mass, $M_d$.
If the accretor is a BH, its own mass may be near or above $10\, M_\odot.$ The value of $M_{tot}$
could be in the range of tens of solar masses.
If the accretor is a NS, its mass 
is likely to be $\sim 1.4\, M_\odot$. However, for both NS and BH accretors, the observed high luminosity  requires a high rate of mass transfer that can be achieved only by donors that are significantly more massive than a NS  and/or highly evolved. 
 The lack of evidence of  
a red luminous giant, is consistent with mass being provided by a high-mass donor. Indeed, as mentioned above, M51-ULS-1 is considered to be a high-mass X-ray binary (HMXB) \citep{2017MNRAS.466.1019S}.   The present-day value of the donor's mass may be a sum of its primordial mass and mass gained during a previous stage of mass transfer from the progenitor of the presently-observed accretor.

The highly luminosity of M51-ULS-1 is driven by a 
high rate of mass infall. An accretion rate of ($\sim 10^{-6}~M_\odot$~yr$^{-1}$ is needed to produce a luminosity of $10^{39}$~erg~s$^{-1}$).  To provide mass at this rate, a star that is not a red supergiant is likely to be close to filling its Roche lobe: $R_d \le (2-3)\, R_L.$. Since the orbital radius is typically just a few times larger than $R_L,$ the Roche-lobe radius, the size of the orbit is linked to the radius of the donor.
 The donor may be a blue
supergiant \citet{Terashima} with radius $ \leq 25\, R_\odot$; it  may be even smaller if the light from the counterpart is blended with light from the accretion disk,
or if the true counterpart is dimmer than the HST-observed emitter. The Roche lobe cannot be more than roughly 2-3 times larger than the donor if mass is to reach the accretor at the high requisite rate.
Putting this all together, we find that  the most probable range for $a_{bin}/R_d$ is $4-10$, with values toward the lower end of the range more likely given the high luminosity.  We therefore
expect the maximum possible size of the binary orbit to be about $3\, AU$,
with the most likely value several times smaller.

\section{Comparing the Transit to Other Light Curve Features}\label{sec:transitfeatures}

\subsection{Accretion-Related Dips}

X-ray light curves exhibit variability of many types.
Flares, long-lasting high and/or low states are observed, as are short-lasting dips that are not transits.  It is therefore important to compare the event we identified with others, in order to determine whether its characteristics set it apart from other dip events. 
We contrast the transit with dip-like behavior found in M51-ULS-1 and in other XRSs in our sample. 

X-ray telescopes not only count the numbers of photons received, but also record the energy of each incoming photon, so we can explore how lower-energy (``soft'') photons behave compared to higher-energy (``hard'') photons\footnote{We have dubbed the higher-energy band ``hard'' (H) and the lower-energy band ``soft'' (S) in keeping with X-ray astronomy convention. See Figure~4 for {\sl Chandra}; Figure~5 for {\sl XMM-Newton}.}. Energy dependence observed during events can provide clues to the cause of the variability.

\begin{figure}
    \centering
    \includegraphics[width=0.90\linewidth]{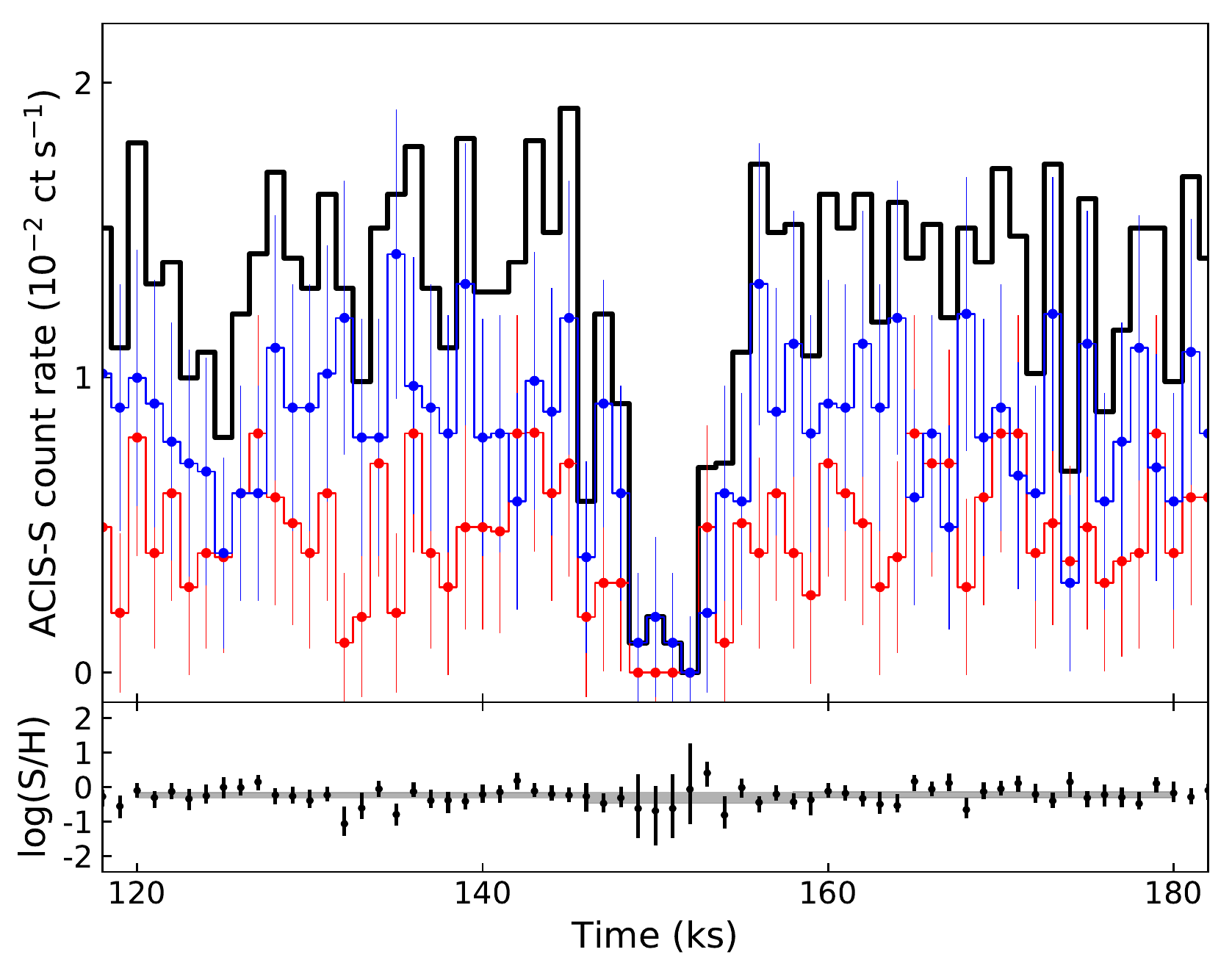}
    \caption{Background-subtracted light curve demonstrating the lack of spectral variation across the transit event in M51-ULS-1.  The {\sl top panel} shows the background-subtracted {\sl Chandra} count rate light curves in the soft ($S$:0.3-0.7\,keV; red histogram) and hard ($H$:0.7-7\,keV; blue histogram) passbands along with the Gehrels-approximated error bars (vertical bars). The solid black line indicates the net rate in the broad ($0.3-7$~keV) passband, with error bars omitted for clarity. The {\sl bottom panel} shows the color hardness ratio ($C=\log{\frac{S}{H}}$) computed using an accurate Poisson model \citep{2006ApJ...652..610P}.  The grey-shaded bands denote the 90\% HPD intervals for counts accumulated over time intervals before, during, and after the eclipse.  The error bars increase in size when the counts decrease during the eclipse, but the spectral hardness shows no evidence of a change. Note the continuity of the grey-shaded bands.}
    \label{fig:eclipse_hardness}
\end{figure}

One case in which energy-dependence should be minimal is during a transition in to or out of eclipse. 
Figure \ref{fig:eclipse_hardness}  demonstrates that despite the sharp drop in intensity during the transit of M51-ULS-1b, 
the spectrum 
shows no evidence of a change. It is clear that the value of the ratio of the numbers of soft (S) to hard (H) photons, quantified by the so-called hardness ratio ($log_{10}(S/H)$) during the transit is consistent with its value out of transit.
This behavior supports the interpretation of the event as a transit. Were the count rate higher, we could quantify the similarity through transit-model fits in different energy bands.

\begin{figure}
    \centering
\includegraphics[width=0.90\linewidth]{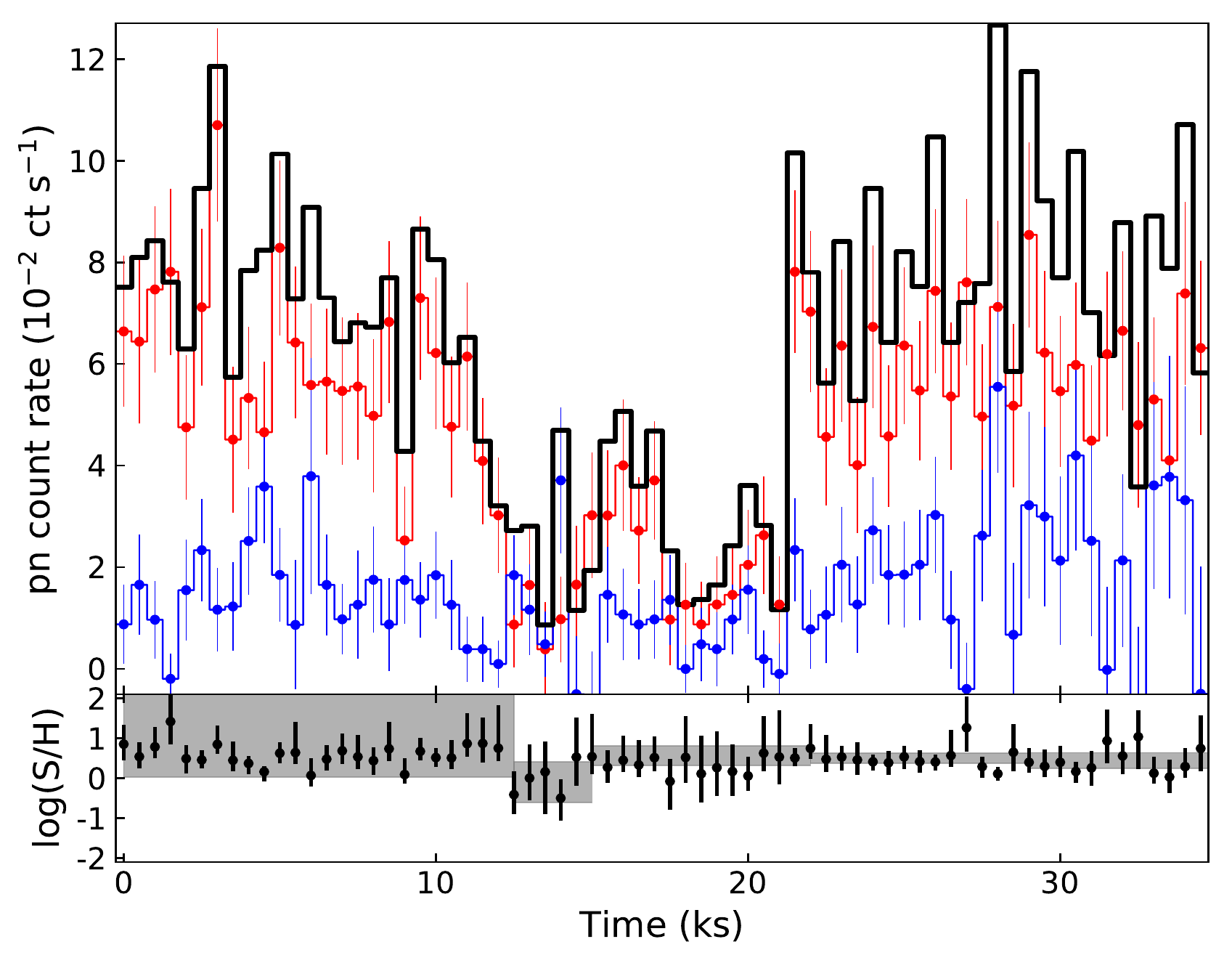}
    \caption{As in Figure \ref{fig:eclipse_hardness}, but for the M\,51-ULS-1 double-dip feature from \textit{XMM-Newton} observation 303420201. Here, the \textit{XMM-Newton} soft and hard passbands are 0.2-0.7\,keV (red histogram) and 0.7-10\,keV (blue histogram), respectively. Unlike the short-duration eclipse, the hardness ratio of M\,51-ULS varies across the duration of the double-dip, suggesting that this event and the one in Figure~3 are caused by different physical processes.}
    \label{fig:m51_doubledip}
\end{figure}

\begin{figure}
    \centering
    \includegraphics[width=0.8\linewidth]{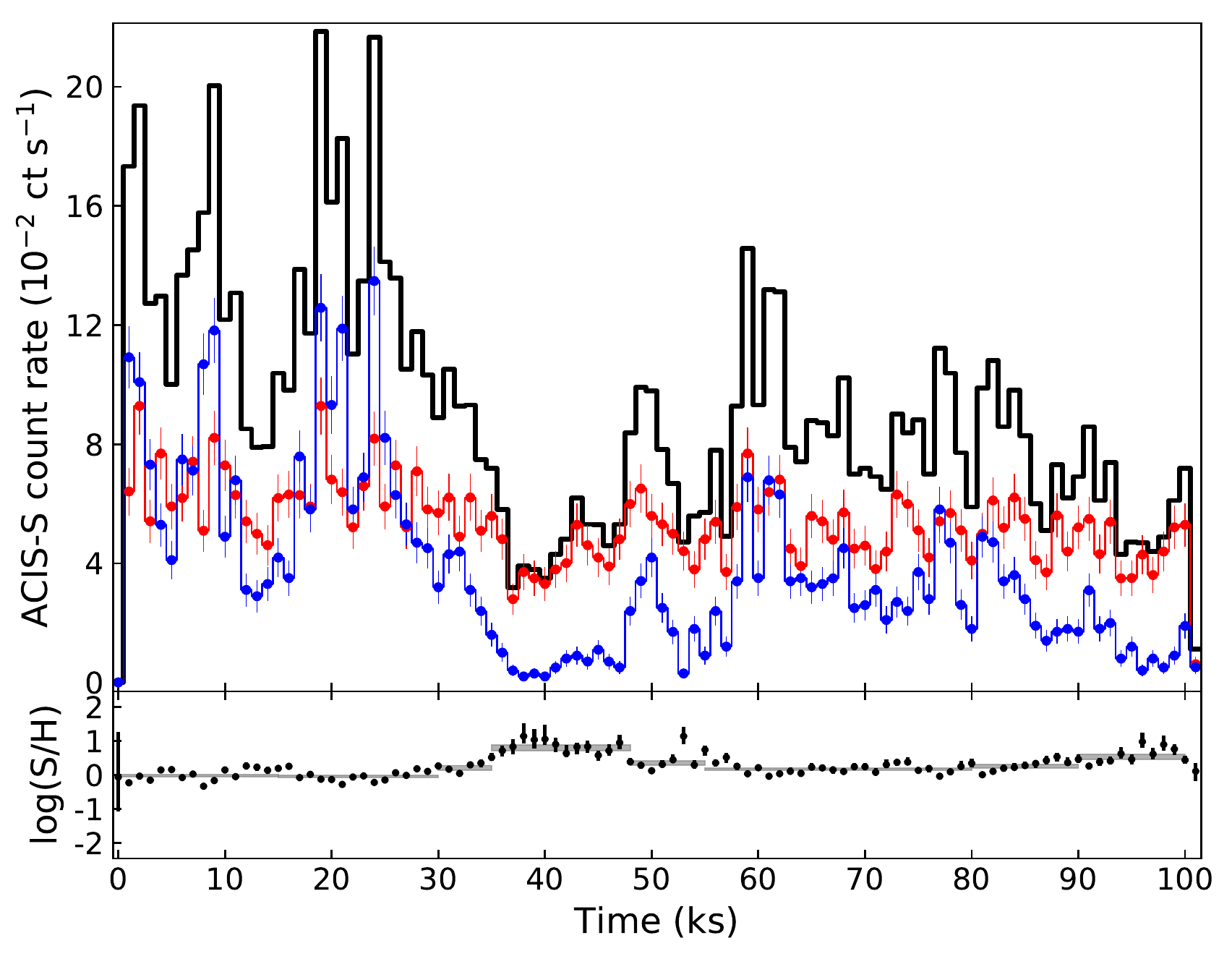}\\
    \includegraphics[width=0.8\linewidth]{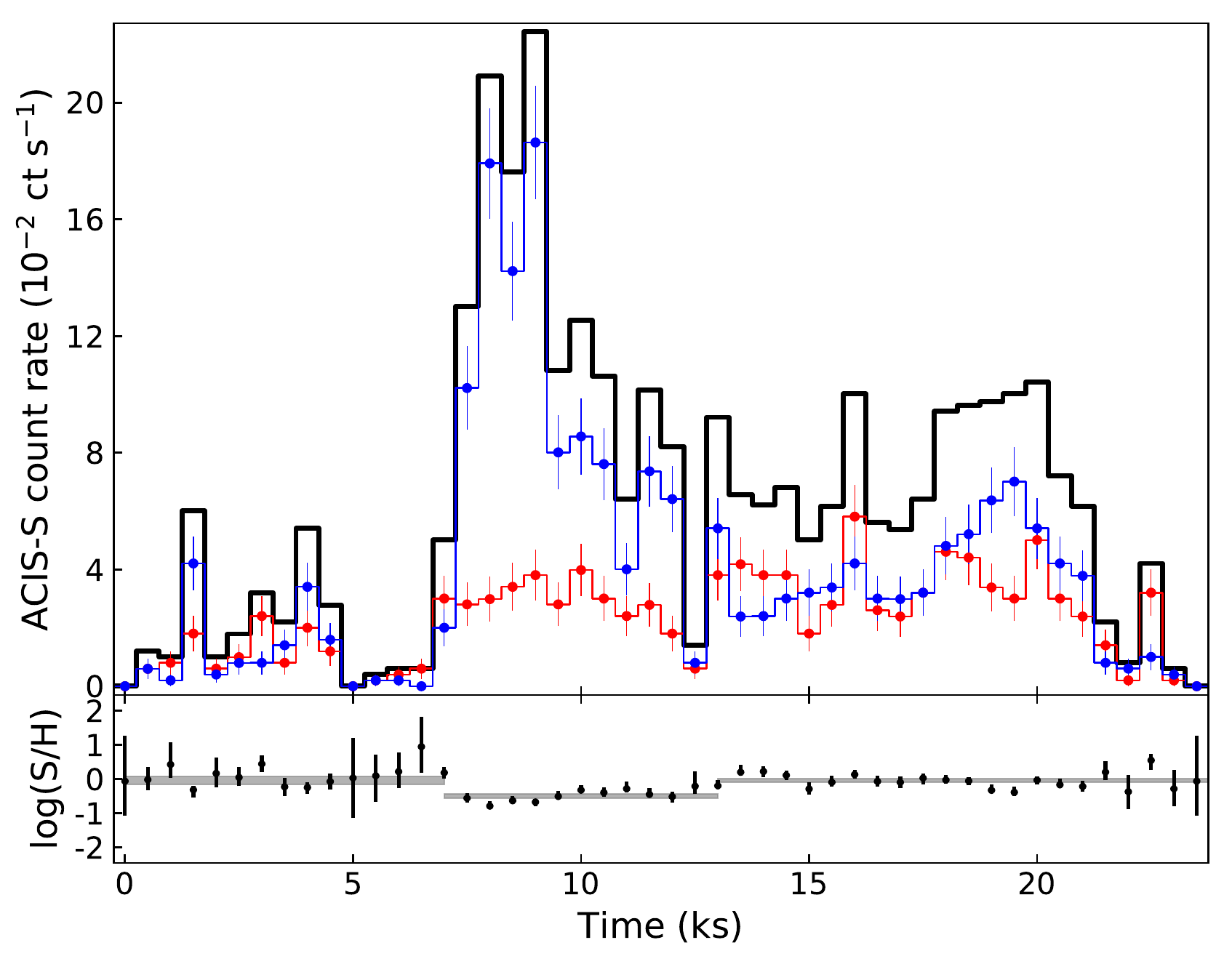}\\
    \includegraphics[width=0.8\linewidth]{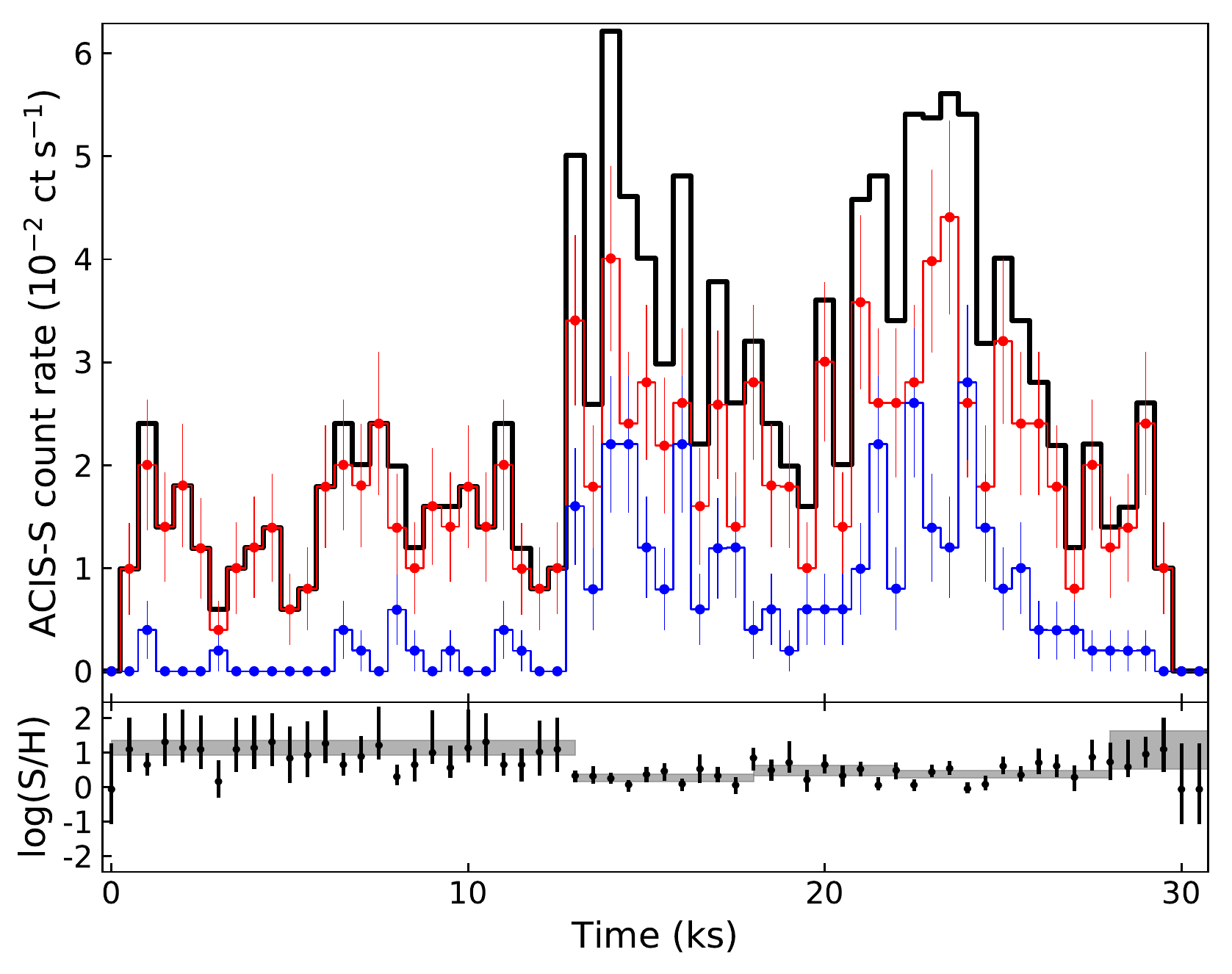}
\caption{As in Figure \ref{fig:eclipse_hardness}, but for M\,101-ULS, with \textit{Chandra} observations 934, 4737 and 5338 (top, mid, bottom, respectively). Each epoch shows strong energy-dependent variability, unlike what is seen for the M\,51 ULS transit event in Figure~\ref{fig:eclipse_hardness}.}
\label{fig:M101_eclipse_lcs}
\end{figure}

Figure~\ref{fig:m51_doubledip} shows a different dip-like event from the light curve of M51-ULS-1. In contrast to \ref{fig:eclipse_hardness}, this dip is not symmetric; the flux doesn't fall to zero; and the hardness ratio changes across the event.  
It serves to illustrate that the transit has distinctive features not typical of
other dipping behavior.

The behavior of the event shown in Figure~\ref{fig:m51_doubledip} suggests that the dip is due to interactions with a high-density feature associated with accretion.
Accretion at high rates,  whether in  
XRBs or young stars \citep{2014AJ....147...82C}, can lead to irregularities in the accretion stream or clumping  of the accreting material.
In XRBs, clumps or portions of the accretion stream may have high enough column density to block X-rays. 
Because, however, these are diffuse structures, they exhibit variations in gas and dust density that translate into different amounts of absorption as they pass near or in front of the XRS.
In such cases, X-rays of different energies exhibit different effects as the density of the material passing in front of the XRS changes. We also note that the passage of material associated with accretion is not generally expected to produce a symmetric light curve dip; furthermore, the irregularity may not be large enough or well centered enough in relation to the XRS to block all of the photons.

This double-dip event not only illustrates ways in which the properties of an absorption-induced event can differ from those of a transit, it also tells us that there are absorbing clouds or streams, likely associated with the accretion disk, along our line of sight to M51-ULS-1. This indicates that our line of sight is aligned with that of the accretion disk of M51-ULS-1. Because the accretion plane is generally aligned with the binary orbital plane, {\sl our line of sight is also aligned with the orbital plane}, suggesting that we may detect eclipses of the XRS by the donor star.

Figure~\ref{fig:M101_eclipse_lcs} shows three events from M101, each of which also provides a clear contrast with the transit event.  The first is an energy-dependent double-dip-like feature similar to the one in Figure~\ref{fig:m51_doubledip}, likely associated with accretion. The two lower panels show transitions from a low-state to a higher state which itself exhibits dips.  These events exhibit clear signs of energy dependence. Thus, the characteristics of these and many other events in the 2624 light curves we studied stand in contrast to the characteristics of the transit.  {\sl The transit in M51-ULS-1 is an {\bf energy-independent} high-low-high transition with a well-define baseline.  It was uniquely selected by our automated search directed at identifying events with the simplest set of features associated with a transit.  It is approximately symmetric, and has a shape typical of transits in which the source and transiting object have comparable size. In \S 5 we will show that it 
is well fit by a transit model.}

\subsection{Intrinsic Variability} 

In addition to variations caused by the passage of matter in front of the XRS, XRBs exhibit a wide range of intrinsic variability.
Intrinsic variations generally show both intensity and spectral changes. 
One particular type of state common to soft XRBs, is an X-ray ``off'' state.  These were first observed in luminous supersoft X-ray sources (SSSs) in the Galaxy and Magellanic Clouds \citet{1996LNP...472..165S}.  For these nearby XRSs, we know that the X-ray flux diminishes to undetectable levels, while the optical flux increases. Although the transitions appear not to have been observed, the behavior is consistent with an expansion, and then later, when the X-ray emission returns,  a contraction of the photosphere \citep{JCGRD}. 
During observations, M51-ULS-1 exhibited at least one clear X-ray off-state (830191401) for which the transition was not observed. We cannot determine whether that off state corresponded to an interval of large photosphere or to the middle portion of an eclipse.  If the system was in eclipse, the eclipse lasted longer than 98~ks.  
As Table~2 shows, there are several additional candidates for X-ray ``off'' states during observations which included no interval of higher count rate. 
We make no assumptions about the nature(s) of the X-ray off states.

Although transitions to off states in SSSs and ULSs have generally not been observed, 
 they are expected to take longer than a ks \citep{JCGRD}. Furthermore, the hardness ratio would change significantly during the transitions into and out of the X-ray off states, in marked contrast to what happens during eclipse. 

\subsection{Stellar Eclipses}

\begin{figure}
    \centering
    \includegraphics[width=0.8\linewidth]{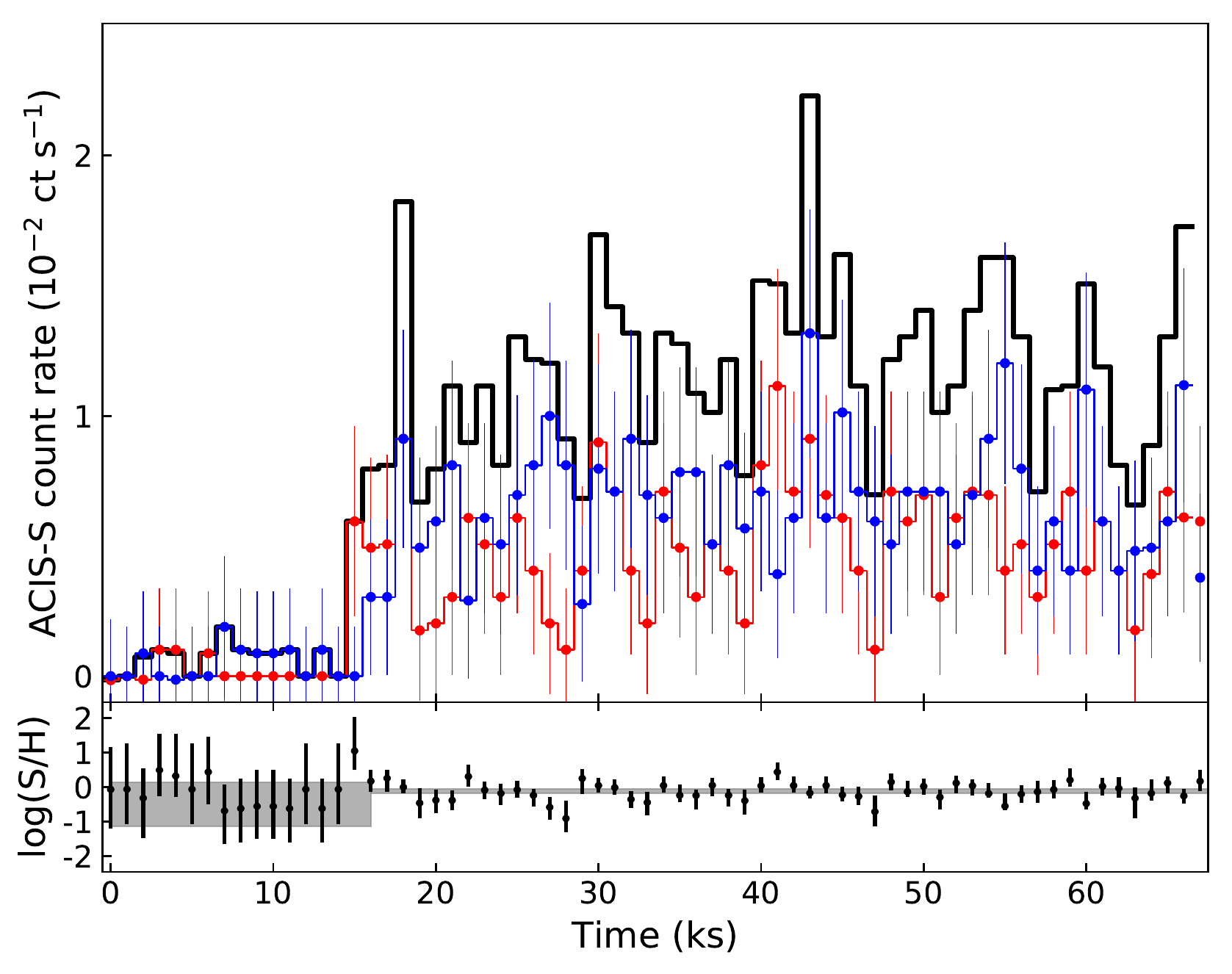}\\
    \includegraphics[width=0.8\linewidth]{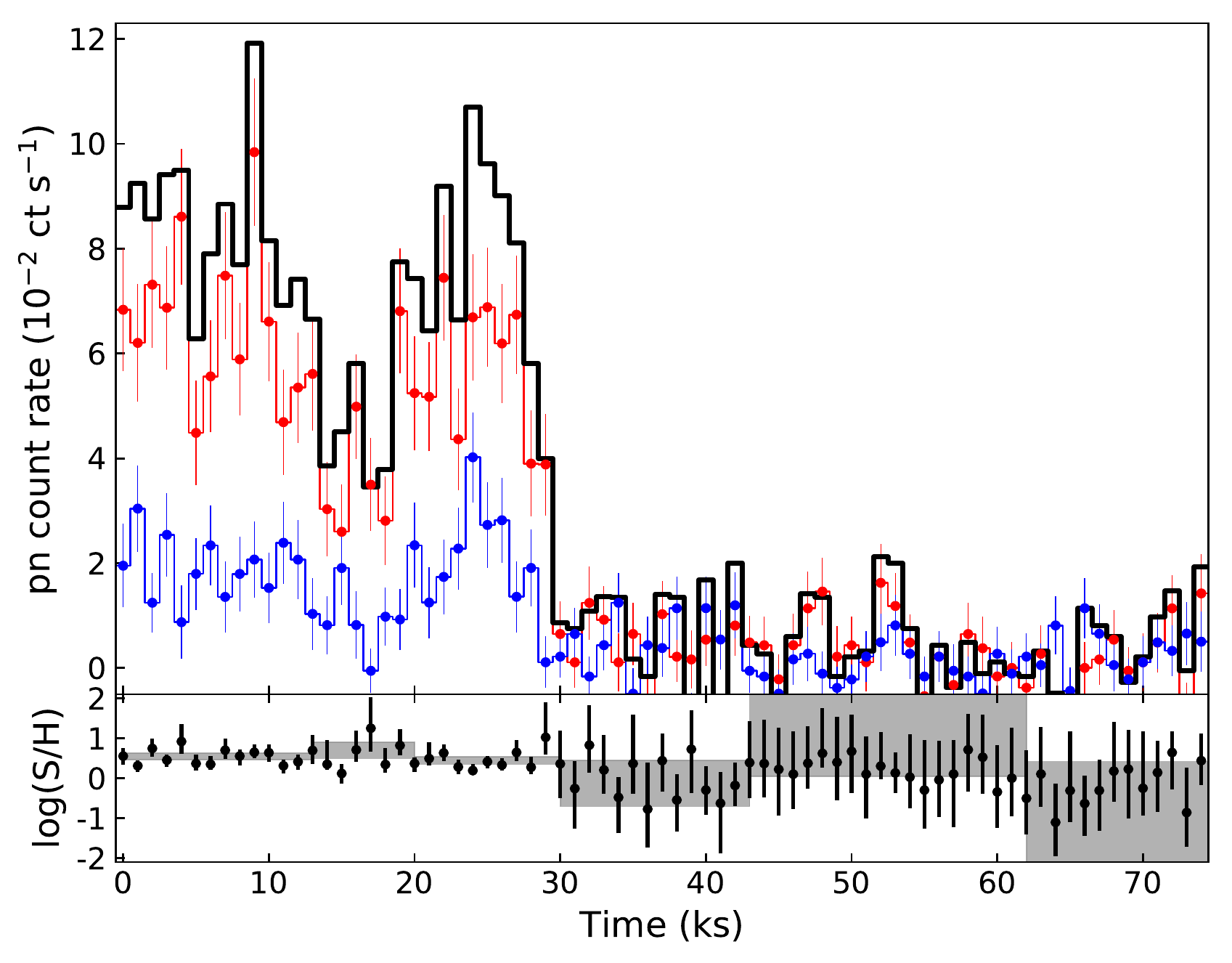}
\caption{As in Figures \ref{fig:eclipse_hardness} and \ref{fig:m51_doubledip}, but for other variable events in M\,51-ULS-1. Left; a long-duration eclipse egress in \textit{Chandra} observation 13815. Right: a long-duration eclipse ingress in \textit{XMM-Newton} observation 824450901. These events are thought to be occultations by the companion star.}
\label{fig:m51_eclipse_lcs}
\end{figure}

In addition to off states extending over an entire observation, there are two observations, one by {\sl Chandra}
and one by {\sl XMM-Newton} in which there was a transition from a low to a high state, and a high to a low state, respectively.
In each case, the transition occurred during an interval of a ks.
The variation shown in the left-hand panel of Figure~\ref{fig:m51_eclipse_lcs}, observed by {\sl Chandra}, exhibits behavior consistent with an egress from an eclipse, presumably by the donor star. The low state is consistent with zero flux.  This state begins prior to the start of the exposure and continues for $15~ks$; the duration of the observed portion of the low state is longer than the full duration of the transit event we present in this paper.
 This event 
 has two characteristics of eclipse: (1) the rapid change from zero flux to a significantly larger count rate, and (2) no change in hardness ratio during the transition.
A change in hardness ratio would likely signal a change in state, whereas during an eclipse the decrease in flux from the harder and softer X-rays occurs at roughly the same time. If the event is an egress from eclipse, the steep rise indicates that, in contrast to the transit, the eclipser is significantly larger than the XRS.

\begin{figure}
    \centering
    \includegraphics[width=0.8\linewidth]{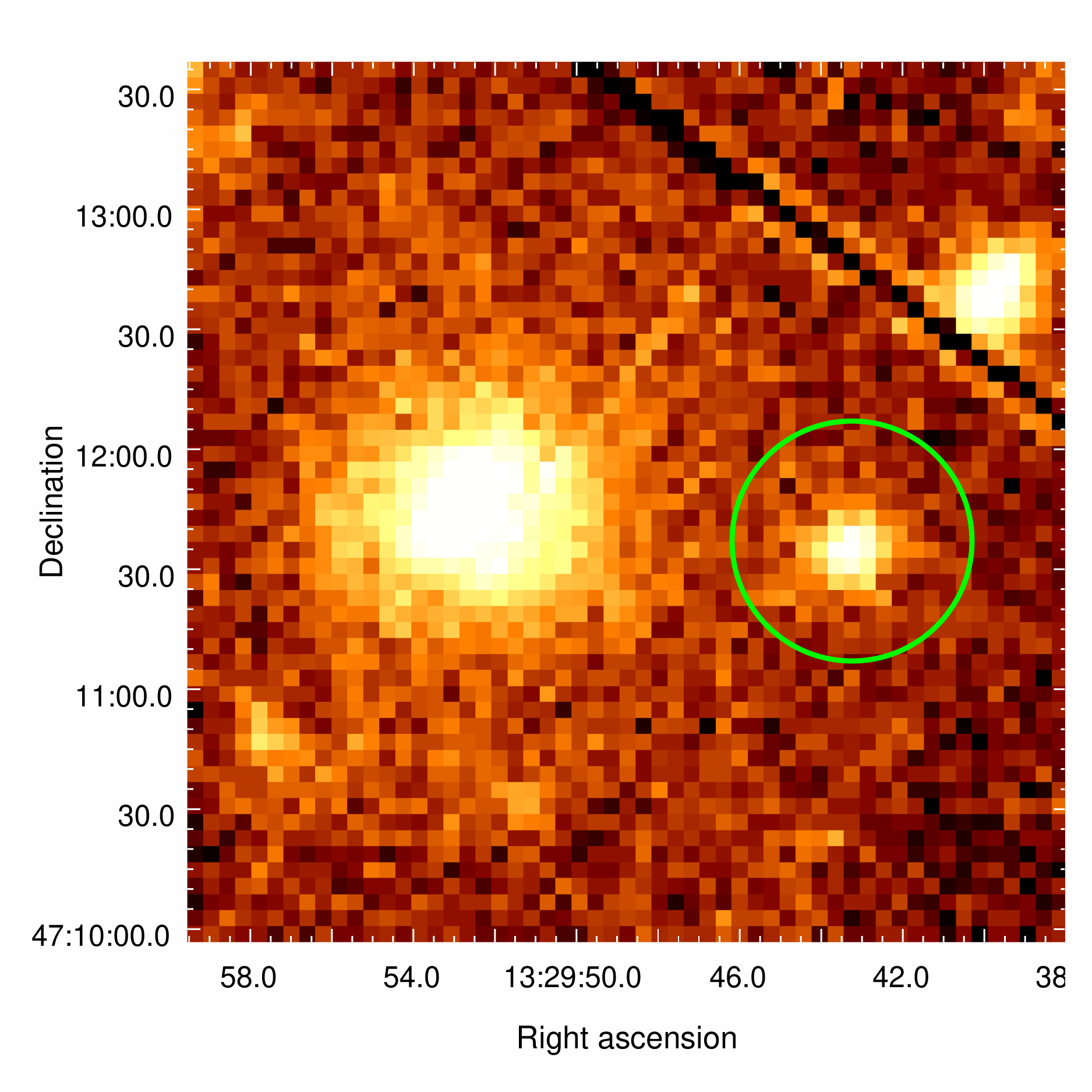}\\
    \includegraphics[width=0.8\linewidth]{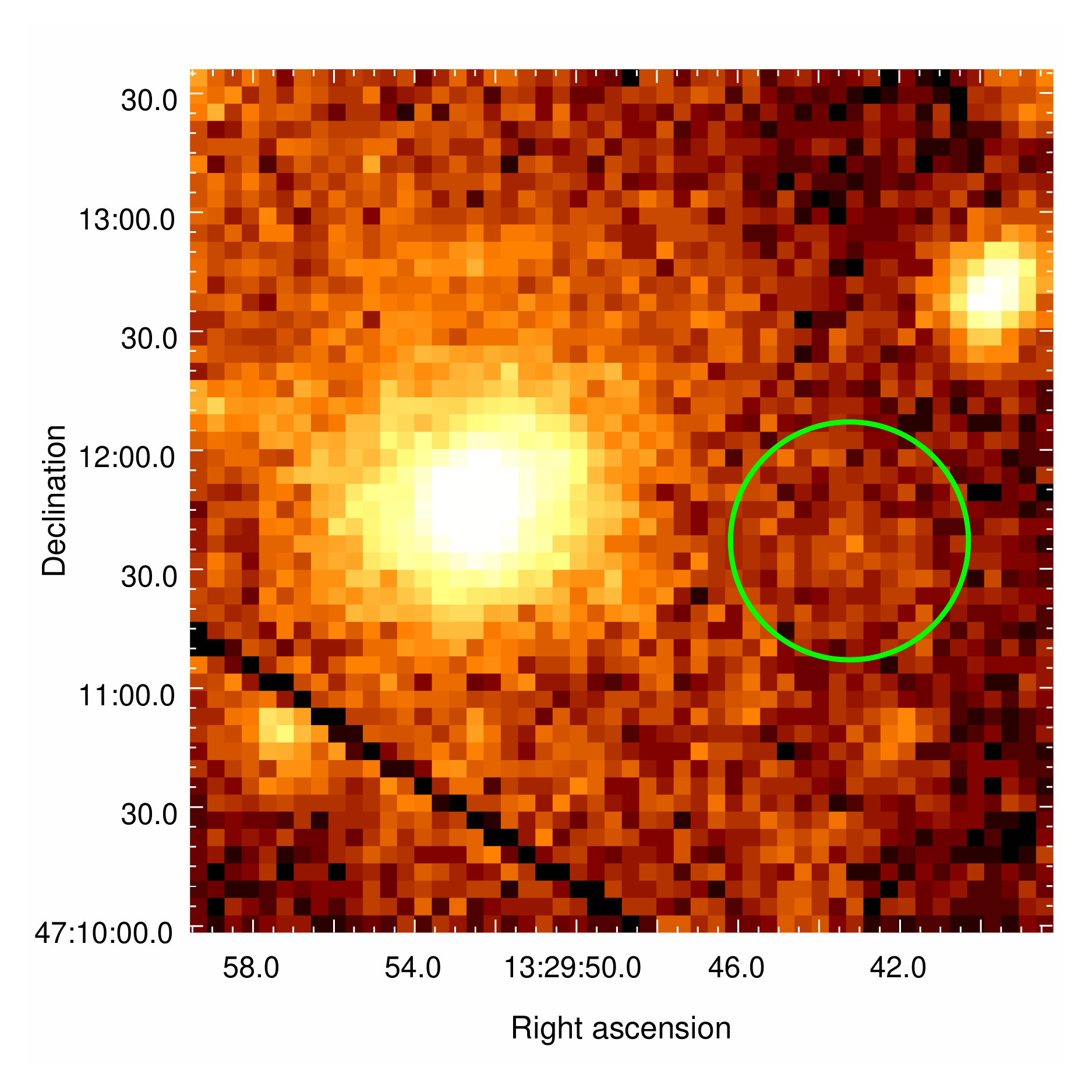}\\
    \includegraphics[width=0.8\linewidth]{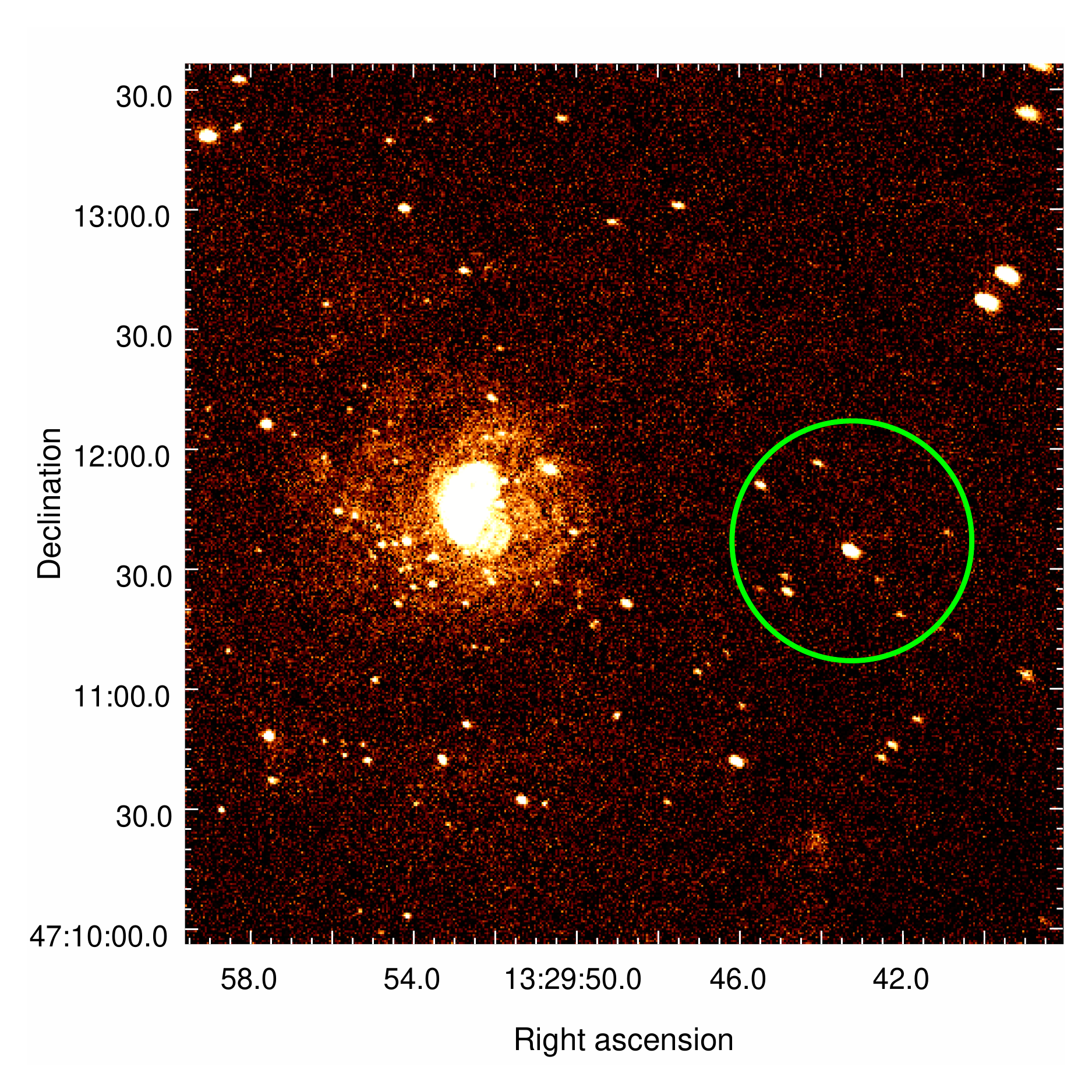}
\caption{Left: \textit{XMM-Newton}/pn 0.2-10\,keV image of the observation 303420201 during which M\,51-ULS-1 appears active. Middle: as in the left panel, but for the portion of \textit{XMM-Newton} observation 824450901 in which the source is in eclipse (i.e., $\sim$30--75\,ks in Figure \ref{fig:m51_eclipse_lcs}, right). Right: stacked 0.3-7\,keV \textit{Chandra}/ACIS image of the same field. The 25$^{\prime\prime}$ radius green circles in each image is comparable to  the \textit{XMM-Newton} extraction region for M\,51-ULS-1. It is clear from the \textit{Chandra} image that a number of nearby point sources are likely contaminating the \textit{XMM-Newton} extraction region for M\,51-ULS-1 and may be causing the residual hard emission seen while the source is in eclipse.}
\label{fig:m51_xmm_field}
\end{figure}

The right-hand panel shows what appears to be an ingress to eclipse, as observed  by {\sl XMM-Newton}. There is a sharp decline as expected for ingress. There is, however, residual emission that shows some spectral variation detectable even during the low state. {\sl XMM-Newton's} point spread function is large, including X-ray emission from sources within a few tens of pc at the distance to M51.
We have examined the Chandra images of M51-ULS-1 and its surroundings and found several (fainter) point-like XRSs and diffuse emission inside the XMM-Newton/EPIC source extraction region (Figure 8). It is therefore reasonable to hypothesize that the low state is a full eclipse, and that the faint residual emission seen during that time interval in the XMM-Newton data comes from those other sources unresolved by XMM-Newton. 
The {\sl Chandra}-observed transition from a low state is an excellent candidate for an eclipse egress, and the {\sl XMM-Newton}-observed transition from a low state is a good candidate for an eclipse ingress. 
The significance of detecting an ingress to or egress from a stellar eclipse is that it tells us that our line of sight is roughly aligned with the orbital plane.  In \S 7 we will show that the probability of an ingress and/or egress occurring during the roughly 1 Ms of observations afforded M51-ULS-1 may be close to unity.

\section{Fits to the Short-Duration Eclipse}
\label{sec:mcmcfits}

\begin{figure}
    \centering
    \includegraphics[width=1.0\linewidth]{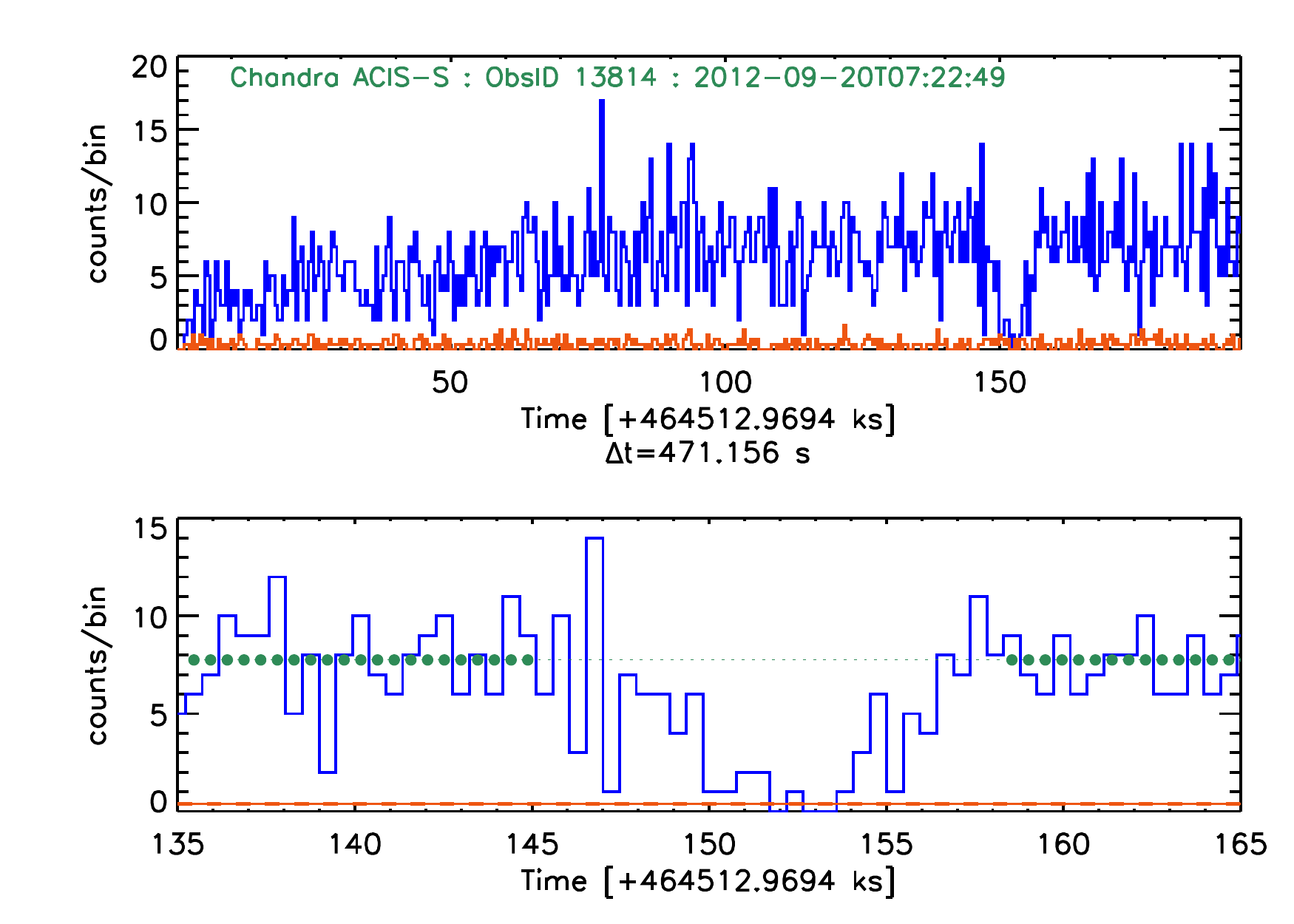}
    \caption{X-ray light curve of the short-duration eclipse, used for the light curve modeling in Section~\ref{sec:mcmcfits}.  {\sl Top panel} shows the light curve of the ULS as the blue histogram over the full duration of the {\sl Chandra} observation.  The contribution of the background, obtained in a source-free region of twice the source area, is scaled and presented as the red histogram.  The light curve is binned at 150$\times$ CCD readout time, so Moire patterns are not expected. {\sl Bottom panel} shows a zoomed in look spanning the eclipse.  The horizontal red line represents the constant background level estimated for the observation, and dashed red lines indicate $\pm{1}\sigma$ errors on the background.  A representative light curve that shows an estimated source count level in the absence of an eclipse is shown as the green dotted line; the level is determined from counts in bins indicated by filled green circles.
    }
    \label{fig:xraylc}
\end{figure}

We model the X-ray light curve (see Figure~\ref{fig:xraylc}) 
using a method that is optimized to analyze low-count X-ray data.  We explicitly use the Poisson likelihood which is appropriate in this regime.  We fit the light curve with a function spanning
the eclipse 
over
the interval from 135~ks to 165~ks.
We represent the XRS as a circular source of radius $R_x$, and the eclipser as an opaque circular disk of radius $R_{ec}=f_{ec} R_x$
We express all distances in units of $R_x$; thus $R_{ec}$ is replaced by $f_{ec}$.
The light curve is defined by five parameters, $\theta=\{\ctsrc,\tmid,\bimpact, f_{ec},\vel\}$, where $\ctsrc$ is the X-ray counts per bin 
outside the eclipse, $\tmid$ is the midpoint of the eclipse, $\bimpact$ is the smallest unsigned distance from the center of the eclipser to the center of the source during the eclipse, and $\vel$ is the velocity at which the eclipser moves across the source.   We use data in the range 135~ks$\leq{t}\leq$165~ks after the start of the {\sl Chandra} observation to construct the light curve; the eclipse occurs approximately between 145~ks and 158~ks.  The events are binned at $\Delta{t}=471.156$~s, corresponding to $150 \times$ the CCD readout duration ({\tt TIMEDEL}=$3.14104$~s).  The velocity is computed in units of $\frac{\rsrc}{\Delta{t}}$, and $\tmid$ in ks starting from the beginning of the observation.  We compute the area of overlap between the foreground object and the X-ray source by considering them as planar circles whose centers ars this area in steradians,
\begin{eqnarray}
    {\rm A}(t) &=& 0 ~~~~{\rm for}~~d(t)>1+\rfront \nonumber \\
    &=& \pi ~~~~{\rm for}~~\max\{1,\rfront\}>d(t)+\min\{1,\rfront\}  \nonumber  \\
    &=& (\alpha_X - \cos{\alpha_X}\,\sin{\alpha_X}) \nonumber \\
    && + \rfront^2 (\alpha_{ec} - \cos{\alpha_{ec}}\,\sin{\alpha_{ec}}) ~~~~{\rm otherwise} \,,
    \label{eq:intersection}
\end{eqnarray}
where $\alpha_X$ and $\alpha_{ec}$ are the angles subtended by the intersecting arcs of the star and the foreground object respectively,
\begin{eqnarray}
    \alpha_X &=& \arccos{\frac{d(t)^2+1-\rfront^2}{2\,d(t)}} \nonumber \\
    \alpha_{ec} &=& \arccos{\frac{d(t)^2+\rfront^2-1}{2\,d(t)\,\rfront}} \,. \nonumber \\
\end{eqnarray}
Note that when the source and eclipser are of the same size, $\rfront=1$, and $\alpha_X$=$\alpha_{ec}$=$\arccos{\frac{d}{2}}$, which results in ${\rm A}$=$\pi$ at $d$=$0$ (complete overlap), and ${\rm A}$=$0$ at $d$=$2$ (complete disassociation).

The model light curve is then computed as
\begin{equation}
    {\rm light~curve}(t) = \ctsrc \cdot \frac{\pi - {\rm A}(t)}{\pi} + background \,,
\end{equation}
where the normalization $\ctsrc$ denotes the expected number of counts in each time bin outside the eclipse, and a time-independent background, scaled from a source-free region of the same observation, is added on.

We carry out the fitting using a Markov Chain Monte Carlo (MCMC) approach \citep{BDA3}.
We use a Metropolis scheme, where new parameter values are drawn based on their current values; we employ as proposal distributions, a Gaussian for $\ctsrc$ and $\tmid$, Uniform for $\bimpact$, and Uniform in log for $\rfront$ and $\vel$.  We further restrict
\begin{eqnarray}
    0< & \ctsrc & < \infty \,, \nonumber \\
    0 \leq & \rfront & \leq 50 \,, \nonumber \\
    0 \leq & \bimpact & \leq 51 \,, \nonumber \\
    10^{-3} \leq & \vel & \leq 10 \,, \nonumber \\
    145 \leq & \tmid & \leq 158 \,.
\end{eqnarray}
We do not explicitly tie together $\rfront$ and $\bimpact$, though the fact that an eclipse is observed naturally requires that $\bimpact<(\rfront+1)$; we expect this correlation to be recovered from the MCMC draws.  We also sample the background level in each iteration from a Gaussian distribution, 
$background \sim N(0.38,0.018^2)$
to account for uncertainty in background determination.  We compute the Poisson likelihood of the observed light curve counts for each realization of a model light curve for the parameter values drawn in that iteration, and accept or reject the parameter draw based on the Metropolis rule (always accept if the likelihood is increased; accept with probability equal to the ratio of the new to old likelihood otherwise).

We first run the MCMC chain for $10^5$ iterations using a starting point of $\theta^{(0)}=\{\ctsrc=7.75,\bimpact=0.5,\rfront=2.0,\vel=0.9,\tmid=150\}$.  We sample 40 different starting points as random deviates from the resulting posterior distributions, and again run $10^5$ iterations for each case.  The first 2000 iterations are discarded as burn-in in each case.  We combine all the iterations after verifying that the chains converge to the same levels for all parameters.  We then construct posterior probability distributions for each parameter as histograms from the MCMC draws after thinning them to the effective sample size ($N_{\rm eff}=\frac{1-\rho}{1+\rho}$, where $\rho$ is the 1-lag correlation) in 5000 iteration increments.

\begin{figure}[H]
    \centering
    \includegraphics[width=1.0\linewidth]{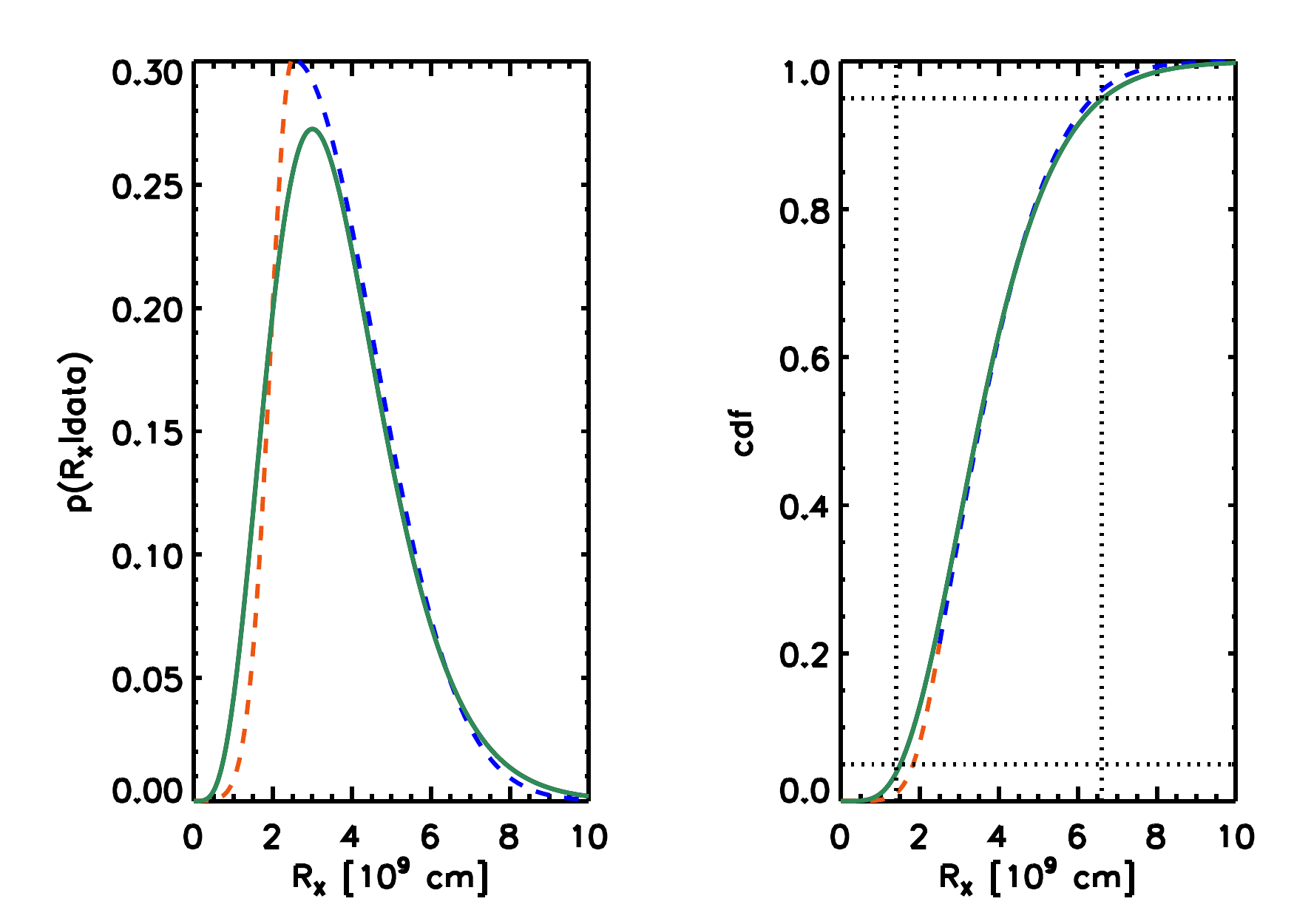}
    \caption{The nominal probability distribution of $\rsrc$.  {\sl Left panel} shows the differential density distributions for the conjoined half-Gaussians (red and blue dashed lines) and for a gamma distribution (green solid line).  {\sl Right panel} shows the cumulative distributions of the two candidate distributions, along with vertical dotted lines indicating the 5\% and 95\% bounds of $\rsrc$, and horizontal dashed lines indicating the corresponding levels.  The gamma distribution matches the bounds better, but has a larger mode.}
    \label{fig:probRx}
\end{figure}

We convert the relative units of $\bimpact$, $\rfront$, and $\vel$ to physical units by convolving the MCMC posterior draws with a representative distribution of $p(\rsrc)$ derived from the X-ray data.  As noted above, the 90\% bounds on $\rsrc$ are asymmetric, at [$-1.1$,$+4.1$]~$\times{10^9}$~cm from the nominal best-fit value of $2.5\times{10^9}$~cm.  This can be represented by half-Gaussians with widths appropriate for the corresponding 90\% bounds (note that for a standard Gaussian distribution, 90\% of the area on one side of the mean is covered at $\pm1.95\sigma$), which are then rescaled to be continuous through the best-fit value which now represents the mode of the pasted Gaussian (see the red and blue dashed curves in left panel of Figure~\ref{fig:probRx}).  However, such rescaling, while it preserves the location of the mode, makes the overall distribution narrower, and the resulting 90\% bounds are no longer consistent with the observed values (see intersections of the dashed red and blue cumulative distribution with the horizontal dotted lines at 5\% and 95\% levels, in the right panel of Figure~\ref{fig:probRx}).  We therefore adopt a Gamma distribution\footnote{The gamma distribution is a highly flexible distribution, defined as $$\gamma(\rsrc,\alpha,\beta)=\frac{\beta^\alpha}{\Gamma(\alpha)} \cdot \rsrc^{\alpha-1}~e^{-\beta\rsrc},$$ where $\alpha$ and $\beta$ are parameters that control the location and shape of the distribution.  The mean=$\frac{\alpha}{\beta}$, variance=$\frac{\alpha}{\beta^2}$, and mode=$\frac{\alpha-1}{\beta}$ are simple functions of the parameters, and conversely, given estimates of the mean, mode, and the variance, we can compute the corresponding parameters.} (solid green lines in Figure~\ref{fig:probRx}) as the representative distribution for $\rsrc$.  Specifically, we choose $\gamma(\rsrc;\alpha=5.36,\beta=1.45)$; the peak here is displaced by $\approx$20\%, but we consider this a better representation of $\rsrc$ because it matches the measured bounds of $\rsrc$ at the 5\% and 95\% levels well.

\begin{figure}
    \centering
    \includegraphics[width=1.0\linewidth]{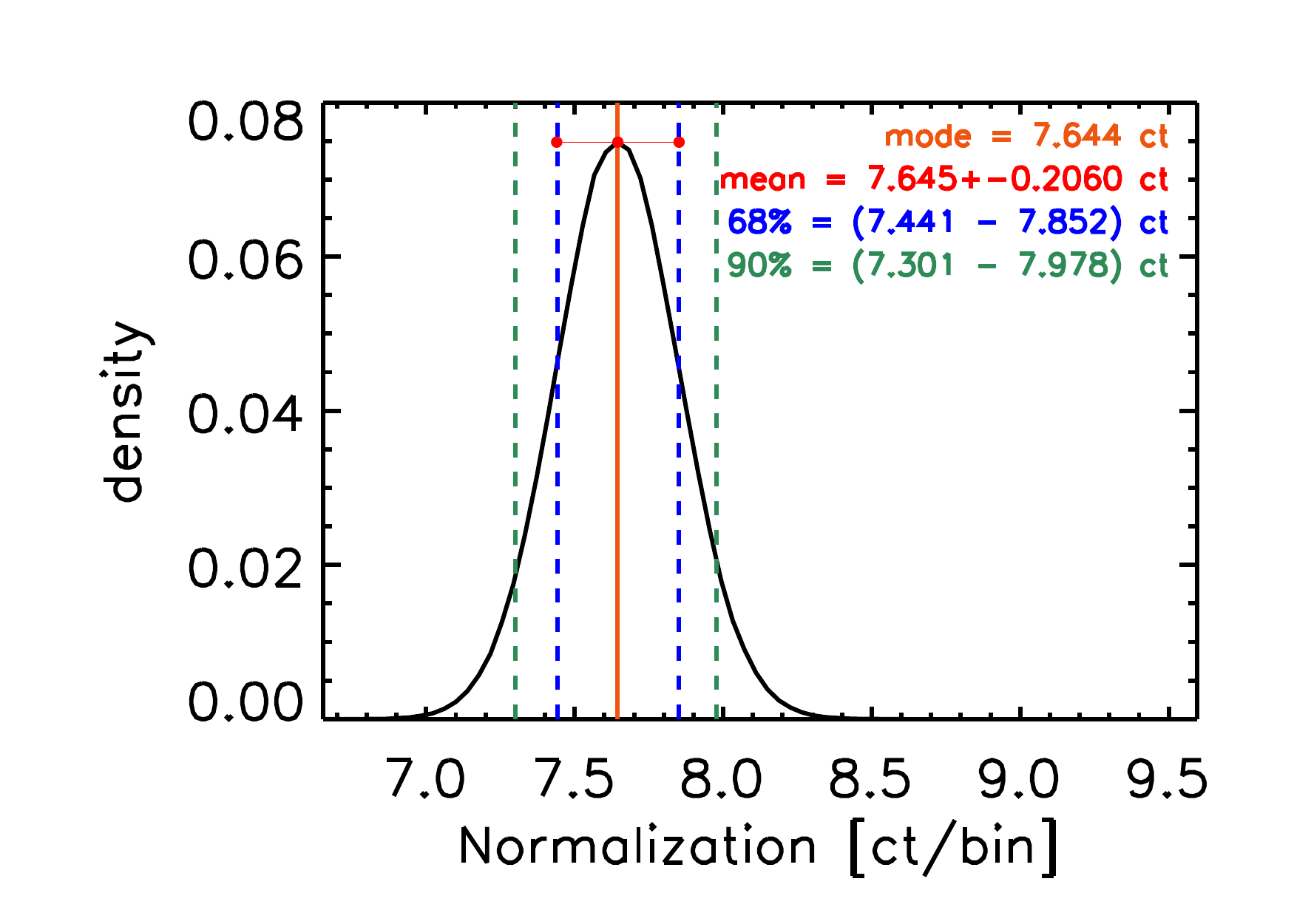}
    \includegraphics[width=1.0\linewidth]{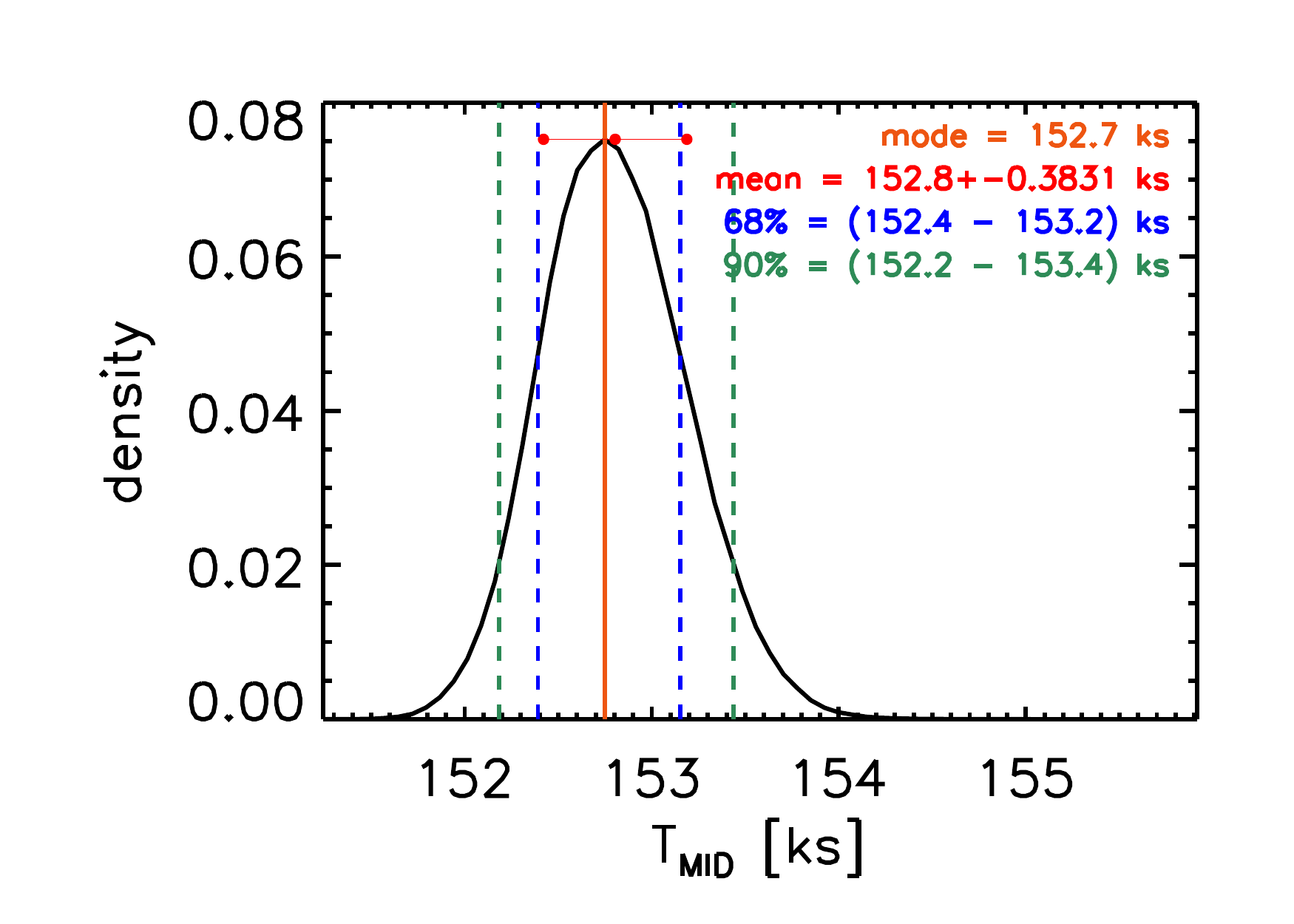}
    \caption{The marginalized posterior density distributions of $\ctsrc$ (top) and $\tmid$ (bottom).  The locations of the mode (vertical solid orange line), the 68\% (blue dashed vertical lines) and 90\% (green dashed vertical lines) highest-posterior density intervals, and the mean and $1\sigma$ errors (horizontal red line placed at the same level as the mode) are marked (see Table~\ref{tab:mcmcresults}), as well as listed at the top right of the panels.
    }
    \label{fig:mcmchisto_normtmid}
\end{figure}

\begin{figure}
    \centering
    \includegraphics[width=0.9\linewidth]{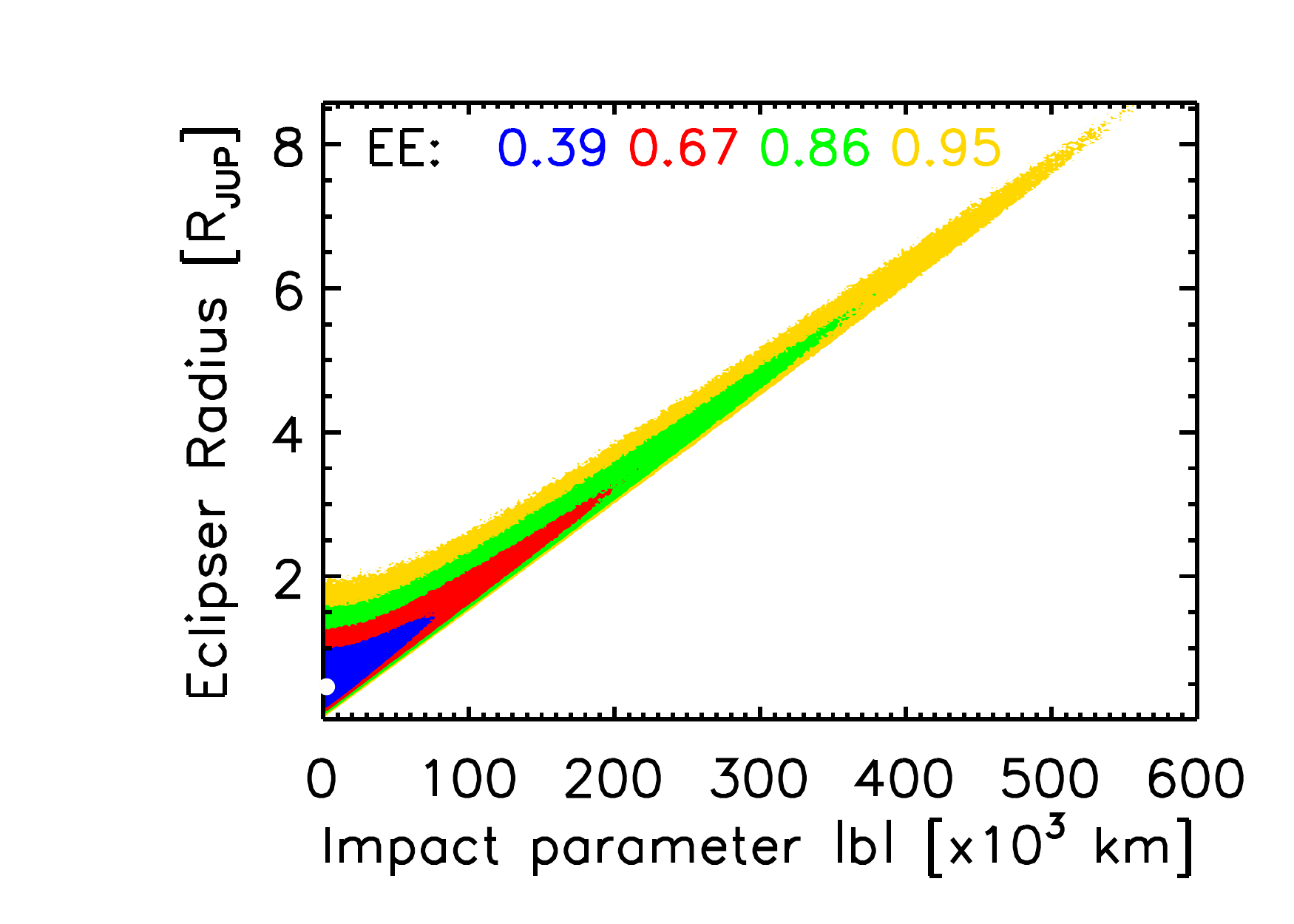} \\
    \includegraphics[width=0.9\linewidth]{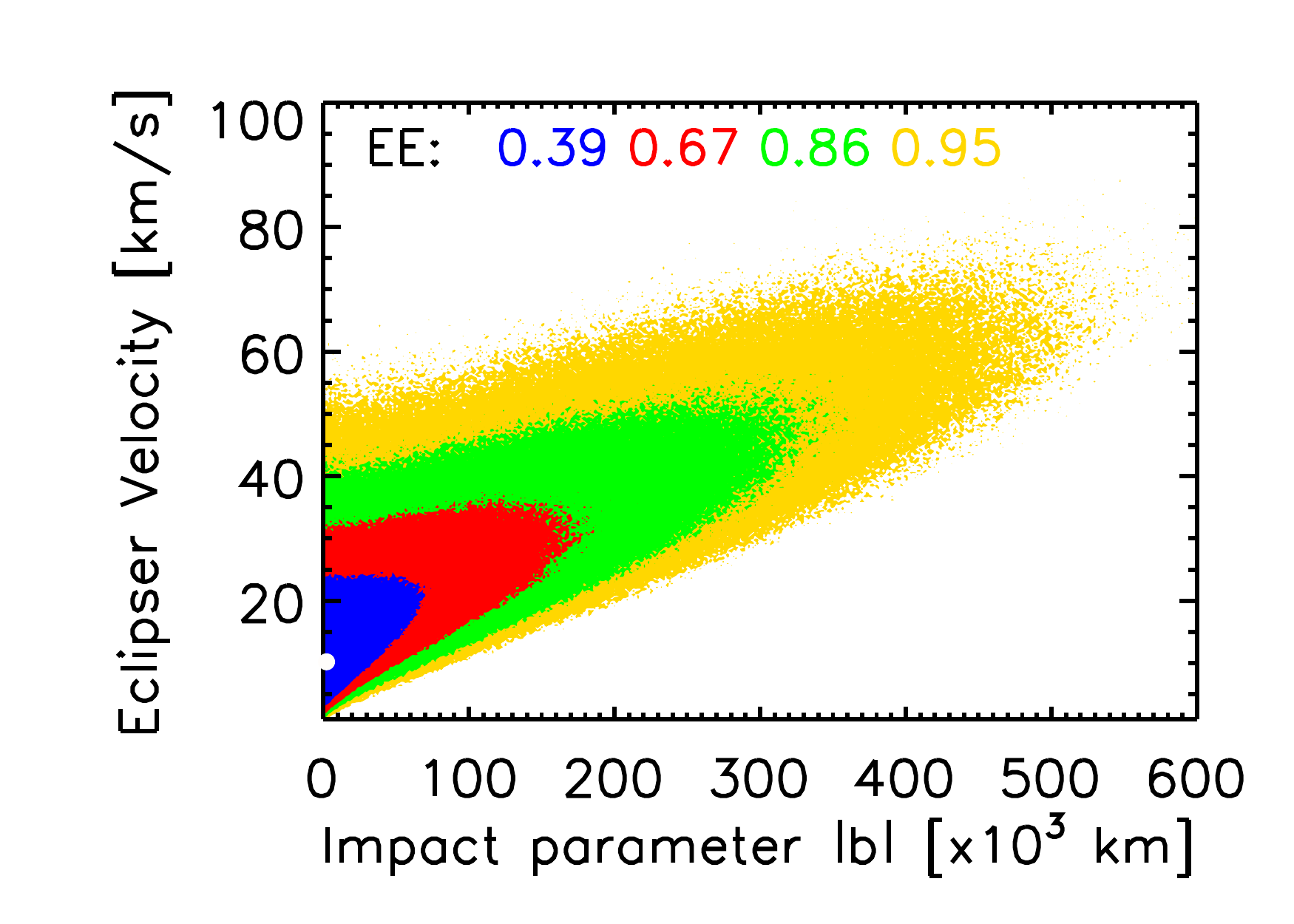} \\
    \includegraphics[width=0.9\linewidth]{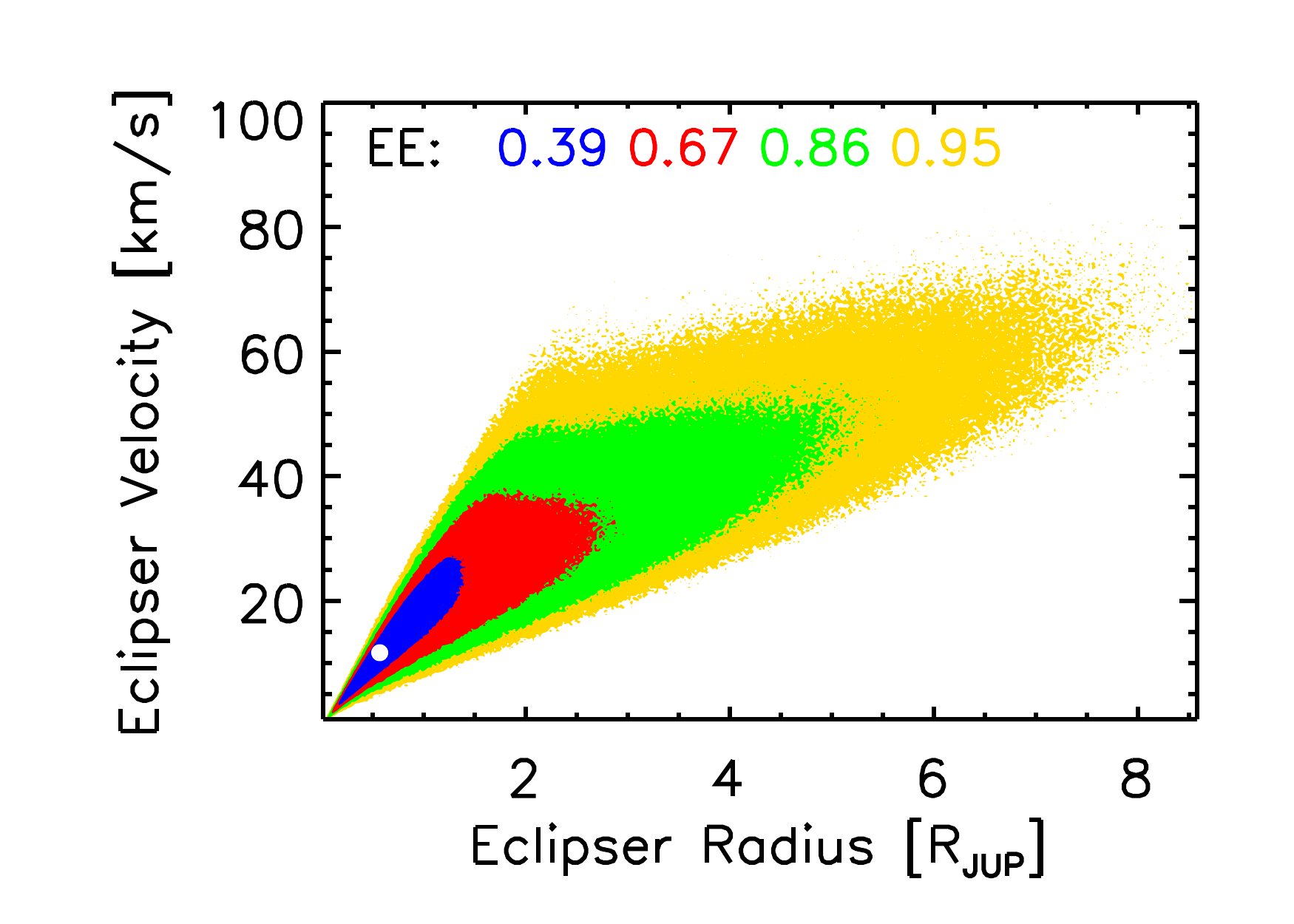}
    \caption{The joint posterior distribution of $(\bimpact,\rfront)$ (top), $(\bimpact,\vel)$, (middle), and $(\rfront,\vel)$ (bottom) are shown.   Each contour plot shows the joint density, marked at enclosed probability regions for 39\% (blue; 2D Gaussian $1\sigma$ equivalent), 67\% (red; $1.5\sigma$), 85\% (green; $2\sigma$), and 95\% (yellow; $2.5\sigma$).  The solid white dot indicates the mode of the distribution.  Note that strong correlations are present between these parameters; see text for discussion.}
    \label{fig:mcmccontour}
\end{figure}

Summaries of parameter values are given in 
Table~\ref{tab:mcmcresults}.
The distributions of the various parameters are shown in Figures~\ref{fig:mcmchisto_normtmid}~and~\ref{fig:mcmccontour}. 
In each case, the location of the mode (vertical solid orange line), the 68\% (blue dashed vertical lines) and 90\% (green dashed vertical lines) highest-posterior density intervals, and the mean and $1\sigma$ errors (horizontal red line situated at the same vertical level as peak of the distribution) are marked.
 Notice that in several cases the distributions are skewed, and traditional estimates like the mean and standard deviation are not useful summaries.  We thus also provide the mode of the distribution and the 90\% highest-posterior density intervals\footnote{These constitute the intervals that encloses the highest values of the posterior probability density, and are consequently the smallest uncertainty intervals that can be set.}  There are also strong correlations present between $\bimpact$, $\rfront$, and $\vel$, as seen from the contour plots of their joint posteriors (constructed without thinning the iterations).  This suggests that the intervals derived from the marginalized 1D posteriors are too coarse, and that narrower intervals may be obtained over smaller ranges.  For instance, $\vel>40$~km~s$^{-1}$ are predominantly obtained when $\rfront>2$~R$_{\rm Jup}$, which itself has a lowered probability of explaining the data.  Thus, the preponderance of the probability suggests that the system is better described with smaller values of $\rfront$ and $\vel$.  Furthermore, notice that $\bimpact$ and $\rfront$ have a large and narrow extension to large values; this can occur essentially because an eclipse can occur for large $\bimpact$ only when $\rfront$ is also large enough to cover the source even at large displacements.  That is, the space of possible models that allow this situation are predominantly driven solely by the depth of the eclipse and not the profile.  This suggests that the number of states that the system can occupy in such configurations is limited, and thus can be described as having low entropy.  This measure is not included in our likelihood, but indicates that smaller values of $\rfront$ and $\bimpact$ are preferred.

\begin{table}
    \centering
    \begin{tabular}{l r r r}
    \hline\hline
    Parameter & Mode & 90\% bounds$^\dag$ & Mean $\pm 1\sigma$\\
    \hline
    $\ctsrc$ [ct~bin$^{-1}$] & $7.6$ & $(7.3, 8.0)$ & $7.6 \pm 0.2$ \\
    $\bimpact$ [km] & $0$ & $(0, 1.8\times{10^5})$ & $(7 \pm 11) \times 10^{4}$ \\
    $\rfront$ [R$_{\rm Jup}$] & $0.74$ & $(0.18, 2.7)$ & $1.4 \pm 1.3$ \\
    $\vel$ [km~s$^{-1}$] & $17.1$ & $(5.1, 56)$ & $30 \pm 20$ \\
    $\tmid$ [ks] & $152.7$ & $(152.2, 153.4)$ & $152.8 \pm 0.4$ \\
    $^\ddag$Eclipse start [ks] & $147.8$ & $(143.9, 151.3)$ & $ 147.4 \pm 2.6$ \\
    $^\ddag$Eclipse duration [ks] & $10.5$ & $(3.1, 17.9)$ & $11 \pm 5$ \\
    \hline
    \multicolumn{4}{l}{$\dag:$ Highest-posterior density bounds} \\
    \multicolumn{4}{l}{$\ddag:$ These are values computed from model parameters, not fitted directly}
    \end{tabular}
    \caption{Results from the MCMC analysis}
    \label{tab:mcmcresults}
\end{table}

\section{The Nature of the Transiting Object}

\begin{figure*}
\includegraphics[width=\linewidth]{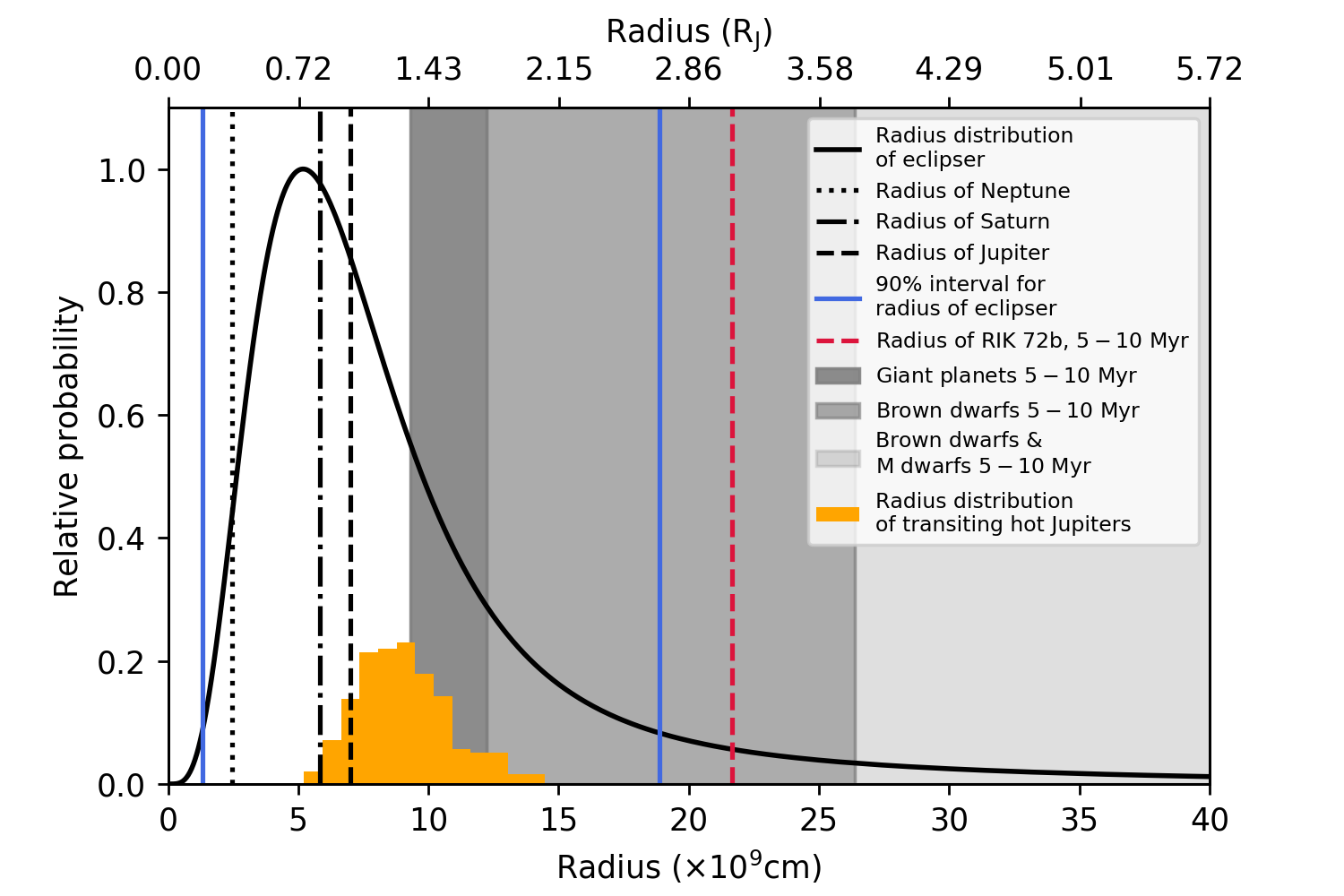} 
\caption{Radius distribution of the eclipser (black solid curve). The lower x-axis is in units of $10^9$ centimeters and the upper x-axis is in units of Jupiter radii ($R_J$). The y-axis represents relative probability for the radius of the eclipser. The left-most (darkest) grey region shows models of giant planets from roughly 0.5 to 10$M_J$ and 5 to 10 Myr old. The contiguous, somewhat lighter grey region corresponds to the range of possible radii for brown dwarfs 5 Myr to 10 Myr;  the lightest grey region on the far right, shows the overlap in radii between brown dwarfs and M dwarfs from 5 to 10 Myr \citep{Baraffe:2003}. The radius distribution of the known transiting hot Jupiter population is shown in orange. The radius of the brown dwarf within the Upper Scorpius association (5-10 Myr), RIK 72b, is shown as the red dashed line.}
\label{fig:radius_ecl_candidates}
\end{figure*}

The size of an object is a powerful indicator of its nature.
Four classes of objects have equilibrium radii in the range consistent with the $90\%$ confidence limits of our model: planets (including rocky planets as well as ice giants and gas giants, e.g. \citep{2019AJ....157..245H});  white dwarfs (WDs);   M dwarfs; (stars with mass less than about $0.5\, M_\odot$); and brown dwarfs.  

WDs have radii that, for different possible WD masses, span
the relevant range of derived values of $R_{ec}.$
WDs are the high-density remnants of stars with initial mass smaller
than roughly $9\, M_\odot$. The radii
of the most massive WDs are comparable to the radius of Earth.  Less
massive WDs have radii of up to a few
times $10^9$~cm.  WDs, however, can only form when the stars that give
rise to them have begun to evolve, generally at ages greater than
$10^8$~yrs.
M51-ULS-1 is almost certainly too young to
be associated with WDs, since most stars that produce WDs will not
yet have begun to evolve. In Appendix B we show that there is an independent reason to eliminate WDs from consideration: in the expected range of distances from the XRS, they would serve as gravitational lenses, increasing the amount of light we receive from the XRS rather than causing a dip.

M-dwarfs have radii that are strongly dependent on age {and irradiation}. For example, a 0.2~$M_\odot$ star has a radius of 13.8~$R_J$ at 1~Myr and 7~$R_J$ at 10~Myr and falls within the 90\% confidence interval at 2~$R_J$ only at ages $\gtrsim$100~Myr. At an estimated age of $10$~Myr, all M dwarfs have significantly larger radii than is estimated for the eclipser. 
Furthermore, the fact that the eclipser is highly irradiated (with typical flux received comparable to that recieved by a hot Jupiter; \S 7) by the XRB  means that it will shrink more slowly toward its equilibrium radius. 
We note in addition  that there are fewer model solutions 
near the upper end of the confidence limits, making the large-$R_{ec}$ solutions less plausible.  

We briefly consider the effects of irradiation on young low-mass objects.
A study in 2003 \citep{Baraffe:2003} produced theoretical predictions for the radii of objects from 0.5$M_J$ to 100$M_J$,  spanning the range from gas giant planets with mass five times that of Saturn to the very lowest mass M dwarf stars. Their calculations explored the effects of irradiation over time for these objects. The environment is at 0.046AU from a host star that has an effective temperature of 6000K. This provides a rough guide to the effects of irradiation.  The radii of these objects monotonically decreases with increasing age so only the oldest objects are found at the smallest radii. For young objects (5 to 10 Myr): giant planets of mass 0.5-13$M_J$ have radii between 1.3-1.8$R_J$, brown dwarfs of mass 13-80$M_J$ have radii between 1.8-5.4$R_J$, and M dwarfs of mass 0.08-0.10$M_\odot$ have radii spanning the range 3.7-5.6$R_J$. 
Figure \ref{fig:radius_ecl_candidates} shows the radius distribution of the eclipser, together with the ranges of sizes predicted by the (sub)stellar models \citep{Baraffe:2003} for brown dwarfs and M dwarfs as functions of age. The radius distribution of roughly 300 transiting hot Jupiters is shown for a comparison to planet-mass objects in an irradiated environment.
On the basis of size, M-dwarfs can be eliminated as candidates for the transiting object.

Brown dwarfs have radii that overlap the upper end of the confidence interval, which as we have noted has fewer solutions and is therefore less likely.   In addition, brown dwarfs are rare relative to planets: several recent studies \citep{Carmichael:2019, Subjak:2019} have addressed this phenomenon, known as the ``brown dwarf desert".  Empirically, a small-radius object is far more likely to be a planet than a brown dwarf.

Thus, although we cannot eliminate the possibility that the transiting object is a brown dwarf, we find that planets are
much more likely.  Planets, unlike brown dwarfs, can have radii across the entire range encompassed by the high-confidence interval, and they are more common companions to stars, as illustrated by Figure  13.

 

\begin{figure*}
\begin{center}
\includegraphics[width=.9\linewidth]{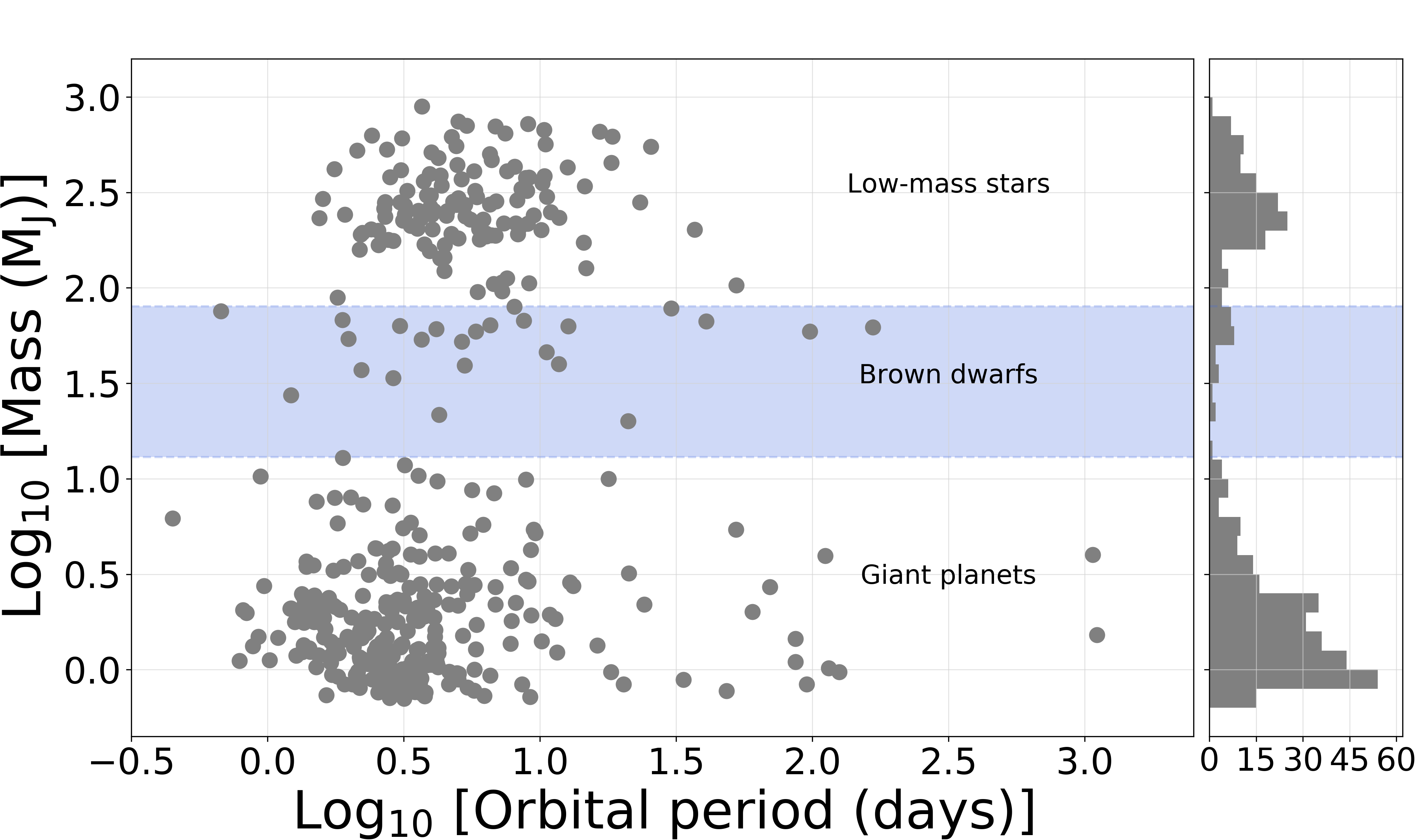}
\caption{Mass-period distribution of a sample of low-mass stellar companions, all known transiting brown dwarfs, and a sample of the transiting giant planet population. The population of brown dwarfs is exaggerated here in that \textit{all} known transiting brown dwarfs are shown while only a sample of the giant planet population (courtesy of http://exoplanet.eu) and a sample of the low-mass stellar companion population \citep{2017A&A.608A.129T} are shown.}
\label{fig:mass_period_dist}
\end{center}
\end{figure*}



\section{The Orbit of the Transiting Mass and its Implications} 
\subsection{The Orbit}

The orbit of the transiting mass determines its position relative to the
XRB and allows us to assess whether it can survive the incoming flux, as well as whether it would have been possible for the candidate planet to survive the evolution of the XRB up until now.  The size of the orbit and the orbital
period also feed into calculations of the probability of transit detection.

For a circumbinary orbit
it is straightforward to determine the value of $a_{pl}$, the
distance of M51-ULS-1b from the binary's center of mass at the
time of transit, since the value of $v_{pl}$ was measured from the short-eclipse fit. 
The most likely value of $v_{pl}$, the mode of the distribution, is 17~km/s, and the $68\%$ uncertainty bounds are at 8~km/s and 34~km/s.

Kepler's law demands that $a_{pl}$ scale as $M_{tot}/v_{pl}^2$, as long as the transiter's mass is much smaller than the mass of the binary.
\begin{equation} 
a_{pl}=45~{\rm AU} \, \Bigg(\frac{M_{tot}}{20\, M_\odot}\Bigg)\, \Bigg(\frac{20\, {\rm km/s}}{v_{pl}}\Bigg)^2 
\end{equation}
The equation above demonstrates that for values of $M_{tot}$ and $v_{pl}$ similar to those we expect, the distance between the candidate planet and the center of mass of the XRB is in the range of tens of AU.
Note that $a_{pl}$ corresponds to the distance at the time of transit.  
In appendix B we show that for orbits wider than a few AU, a WD would
produce a lensing event rather than a dip in flux.  This, in addition to the young age of the system, eliminates the possibility that the transiting mass could be a WD.

With an orbital radius on the order of tens of AUs,  M51-ULS-1b orbits both components of the XRB.  However, even if the combination of $M_{tot}$ and $v_{pl}$ is such that the value of $a_{pl}$ is only on the order of several AU, M51-ULS-1b's orbit is still almost certainly circumbinary, since the binary's orbital radius is likely to be a few times smaller than the maximum value of $3$~AU. 

We now consider the requirement of orbital stability. The semimajor axis of the candidate planet's orbit and the binary's orbital radius must be larger than roughly $3$.  This suggests that the planet-candidate's orbit forms a hierarchical system with the XRB; a  value of $a_{pl}$ in the range of several AU or higher is consistent with the XRB's properties and the condition of orbital stability.

\subsection{Incident Flux and the Survival of M51-ULS-1b}

The XRS is highly luminous. 
We compute the ratio  of the flux incident on the planet candidate to the flux incident on Earth from the Sun. The luminosity of the Sun is $4\times 10^{33}$~erg~s$^{-1}$, and we take the luminosity of M51-ULS-1 to be $10^6$ times larger.
\begin{equation}
    \frac{\cal F}{{\cal F}_{\oplus}} 
    = 
    490\, \Bigg(\frac{45\, {\rm AU}}{a_{pl}}\Bigg)^2 
    = 156 \, \Bigg(\frac{80\, {\rm AU}}{a_{pl}}\Bigg)^2 
\end{equation}
This is similar to the flux incident on a planet orbiting a solar-luminosity star at 0.05~AU.  Gas giants found in such orbits are referred to
as ``hot Jupiters''.  
 
 The high effective temperature of the XRS ($\sim 10^6$~K) means that it is not only a copious emitter of X-rays, but also that a large fraction of the radiation it emits  is  highly ionizing. 
 Such radiation can lead to the loss of the planetary atmosphere.
 Although highly luminous systems like M51-ULS-1 have not yet been considered as planetary hosts, an analogous case has been studied.  Specifically, main-sequence Sun-like binaries whose components are close enough to interact tidally have been studied as hosts 
 for circumbinary planets
 \citep{2014A&A...570A..50S}. The stars in such systems have active coronospheres, and are therefore more luminous in X-rays than they would be had they been isolated.
Calculations conducted for several such real systems have explored the range of parameters consistent with planetary survival. At the distance we have estimated for M51-ULS-1b from its XRS,  its atmosphere can survive the presently observed X-ray active phase of M51-ULS-1.

  At optical and infrared wavelengths the dominant source of flux may be the donor star, although 
the magnitude of the HST-discovered counterpart suggests that the donor does not have a higher bolometric luminosity than the XRS. Thus the discussion above will not be significantly altered by including the effects of the donor star.

In summary, a candidate planet in the orbital range we derive for  M51-ULS-1b can survive its present-day conditions.  This is in contrast to what would be expected
for gas or ice giants in close orbits with M51-ULS-1, which would have their envelopes destroyed on relatively short time scales. Of course M51-ULS-1b is influenced  by the incident radiation.  In analogy to close-orbit exoplanets it would experience bloating, having a radius somewhat larger than expected for an object of the same mass in a region without the large amount of incident flux.  Bloating would also affect brown dwarfs and low mass stars in this environment.
See, e.g., \citep{2019MNRAS.488.3067H}.

\subsection{Feasibility of Wide Orbits}

We know that the existence of planets in wide orbits is plausible, because planets with orbits having semimajor axes in the range of tens and hundreds of AU are common among Galactic exoplanets. Direct imaging has led to the discovery of 15 confirmed exoplanets with estimated mass smaller than $13\, M_J$ and semimajor axes between $10$~AU and $100$~AU;  similarly, 12 exoplanets have semimajor axes wider than $100$~AU ({\sl exoplanet.eu}; 9 July 2020). 
There is also a case of a planet in a $23$~AU orbit about a former XRB in M4 \citep{2000ApJ...528..336F}.

Even without a detailed evolutionary model, we know that the binary
M51-ULS-1 had an interesting history.  Here we discuss key elements of that history and show that a wide-orbit planet could survive. 
M51-ULS-1 experienced an earlier phase of activity during which the star that evolved into today's compact accretor was active. This star could have transferred mass to its companion.  Because, however, the companion was not compact, less accretion energy would have been released per unit mass than is released today. The evolution of the most massive star  would, however, have had consequences for a circumbinary planet.  The evolving star would become more luminous and larger.  It would also shed mass through winds, which would tend to make the planetary orbit wider. If a significant amount of mass was ejected in the orbital plane, the planet's orbit would likely have been driven toward the midplane.  It is important to note that, even if the first-evolved star reached giant dimensions, a planet in an orbit that was initially several AU wide would be able to survive and would likely be pushed into a wider orbit. If the present-day compact object is a BH,  the formation event may have been a ``failed supernova'' during which little mass was lost. If instead the present-day compact object is a NS, significant mass may have been lost through a supernova, and there may also have been a ``kick''.  Nevertheless, the presence of the massive star that is today's donor would have allowed the system to survive and would have moderated the speed of NS's natal the kick. Just as it is possible for a binary to survive a supernova explosion, it is also possbile for wide-orbit planets to stay bound.

The bottom line is that the wide orbit we derive is consistent with the existence and survival of the planet, both in the presently-observed binary and through the possible evolution of the primordial binary, even though not every planet hosted by the XRB will have the same fate.

 \section{Galactic Populations of Planets}
 
 \subsection{The Number of Planets in our Sample}
 
 What is the probability of observing a transit by a small object in our data set?  By answering this question we can use our detection of M51-ULS-1b to estimate how many planet-size objects are likely to be orbiting the XRBs whose observations comprise our data set.  


In Appendix~C we derive an expression for ${\mathbb P}_{trans},$
the probability of detecting a transit. 
\begin{equation}
 {\mathbb P}_{transit}=7.8\times 10^{-5} \, g\, 
 \frac{1}{\alpha}
 \Big[\frac{T_{obs}}{{\mathrm Ms}}\Big]
 \Big[\frac{45 \, {\mathrm AU}}{a_{pl}}\Big]^\frac{3}{2}
            \Big[\frac{M_{tot}}{20\, M_\odot}\Big]^\frac{1}{2} 
\end{equation}
New quantities of the right-hand side are: $T_{obs}$, the total time duration of the X-ray observations; $\alpha$, a parameter whose value is ${\cal O}(1);$ and $g$.  The value of $g$ is unity if the planetary orbit and the binary orbit are coplanar; otherwise it is smaller. Coplanarity appears to hold, at least approximately in M51-ULS-1/M51-ULS-1b, suggesting that $g$ is likely to have a value larger than $0.01-0.1$.
 
 We can use ${\mathbb P}_{trans}$ to estimate the number of wide-orbit planets likely to be present in the sample of XRBs we studied.
 There were $238$~XRSs satisfying our selection criteria in our sample of three galaxies, each with a total exposure time comparable to $1$~Ms. 
Not every XRS detected within the area covered by a galaxy is an XRB.  \footnote{Supermassive BHs at galaxy centers may emit X-rays, and supernova remnants can be bright XRSs. Some XRSs may be distant quasars or nearby stars. Many of the latter can be identified by crossmatching with data from other surveys, such as Gaia\citep{Gaia}.}

The probability of detecting a transit by a planet candidate  in our survey is estimated by multiplying the probability ${\mathbb P}_{trans}$ by $200 \times (N_{xrb}/200)$. Since we observed a single transit, the number of wide-orbit planet-size objects around the XRBs in our sample can be estimated to be:
\begin{equation}
N_{pl}= \frac{64\, \alpha}{g}\, 
\Bigg(\frac{a_{pl}}{45 \, {\mathrm AU}}\Bigg)^\frac{3}{2}
\Bigg(\frac{20\, M_\odot}{M_{tot}}\Bigg)^\frac{1}{2}
\Bigg(\frac{200}{N_{xrb}}\Bigg)
\Bigg(\frac{\mathrm Ms}{T_{obs}}\Bigg)
\end{equation}
Our detection of a planetary transit may therefore signal the presence of roughly $64/g$  substellar objects in wide orbits around the XRBs in our sample.  Note that, because $g$ is almost certainly smaller than unity, the number of small-radius objects orbiting the XRBs in our sample could be even larger than several dozen.  

 Some
XRBs may be more likely to host planets than others,
but more investigations are required to determine relative populations. 
Furthermore, our search may not have discovered smaller objects in equally wide orbits, or even larger objects in closer orbits.


 \subsection{Prospects for Future Observations}

 There is no reason to suggest that the data sets we employed are extraordinary.  Were we to examine a set of XRSs drawn from similar extragalactic populations we would expect a similar result. We therefore examined archived data to determine how many independent and roughly
 equivalent studies could be conducted, to explore the prospects for future discoveries.   Both {\sl XMM-Newton} and {\sl Chandra} data are available for this purpose.  {\sl XMM-Newton} provides the advantage of a la
 rger effective area, yielding higher count rates. {\sl Chandra}'s low   noise and superior spatial resolution, mean that there is little or no confusion, even in relatively crowded fields. Archived and new data from both observatories can discover short-duration transits.

 A search of the {\sl Chandra} archive found that at least 7 galaxies have been observed for $750-1500$~ks, and $13$ others for $250$~ks to $500$~ks.  Two of the best observed galaxies, M31 and M33,  are members of the Local Group, where  sources tens to a hundred times less luminous than the ones we have studied provide enough photons to allow the detection of short transits. Data from dozens of other galaxies with shorter observations are also useful. {\sl XMM-Newton's} archives are comparably rich.

 In short, the archives contain enough data to conduct surveys comparable to ours more than ten times over. We therefore anticipate the discovery of more than a dozen additional extragalactic candidate planets in wide orbits.  Furthermore, additional data from external galaxies is collected every year. 
 Below we discuss how existing data can additionally be used to search for planets with closer orbits and also for planets orbiting dimmer XRBs.
 
 The reason external galaxies are good places to hunt for planets is that the field of view of today's X-ray telescopes encompass a large fraction of the bright portions of galaxies at distances larger than $6-7$~Mpc. This means that a single observations can collect counts from dozens to hundreds of XRSs.  
 As we consider galaxies nearer to us, the advantage of a broad field of view is diminished. There is nevertheless a significant advantage to be gained, because, for example at the distance to M31 we collect
 $\sim 100$ times as many counts as we would from the same XRS in a galaxy at $8$~Mpc.  In addition, for bright sources, this makes us sensitive to shorter-lived deviations from baseline. Thus, small planets in orbits like that of M51-ULS-1b can be detected, and planets in closer orbits can be detected as well.  Furthermore, since the numbers of XRSs at lower luminosities is larger than the number of high-luminosity sources, planet searches can be conducted on the much larger populations of dimmer XRBs. For example, the central region of M31 contains roughly 400 XRSs with total observing times larger than 0.5~Ms, most with luminosities between $10^{36}~{\rm erg~s}^{-1}$ and $10^{38}~{\rm erg~s}^{-1}$. At lower luminosities substellar objects may be able to survive for longer times in smaller orbits.  We will be able to either discover such planets or place meaningful limits on their existence.
 
 Finally, the closest XRSs to us are in our own Galaxy, where even light curves of WDs that accrete from close companions at low rates (cataclysmic variables), with luminosities on the order of $10^{31}$~erg~s$^{-1}$ can be examined for evidence of transits. Unless, however, the target is a cluster or other crowded field, only one XRS may be in a single field of view.  Furthermore, long exposures are available for smaller numbers of XRSs than in external galaxies.  Nevertheless, some XRSs  have had excellent time  coverage by, for example, the past X-ray mission, RXTE or with the current NICER mission. 
 In addition, new missions, such as ATHENA and the proposed Lynx mission, will increase the X-ray count rates significantly, making it possible to discover more planets in all of the environments considered above.
 
 \subsection{Conclusion}
 
 It is worth noting that it has been possible for us to find something as new as an X-ray transit due to a candidate planet, simply because we were looking for it.  XRBs are so variable, and  dips due  to absorption are so ubiquitous, that transit signatures are not readily recognized\footnote{The signal we report on here with the full participation of all coauthors was originally misidentified by two of us \citep{2016MNRAS.456.1859U}.}  Yet, because planets have been found in all environments that have been searched for them,  it is reasonable to look for signs of planets in XRBs.  Once the results from successful searches are known, new discoveries are likely to emerge from a variety of research groups who may take new looks at interesting light curve features.

Our discovery of a single transit will lead to more detailed studies of planets and other low-radius objects in external galaxies.
An equally thorough study of independent data sets will be important to  develop better statistics. It is within the reach of the present generation of X-ray telescopes to develop information about the population of planets orbiting XRBs and for future generations of instruments to develop a comprehensive view.

The discovery of M51-ULS-1b has established that external galaxies host candidate planets.  It also demonstrates that 
   the study of X-ray transits can reveal the presence of otherwise invisible systems, which will also include brown dwarfs and low-mass stars.\footnote{Our method is also capable of discovering them. That a planet-size object was the first discovery may simply reflect a larger population of circumbinary planets than brown dwarfs or M dwarfs.} Discovering and studying extragalactic planets and other small objects in external galaxies can establish connections and contrasts with the Sun's environment in the Milky Way, provide insight into the mutual evolution of stellar and binary orbits, and expand the realm within which we can search for extraterrestrial life. Extending the search will expand the scope of what we can say about our place in the universe.  
 \newpage

{\bf Data availability:} 
The {\sl Chandra} and {\sl XMM-Newton} data
that support the findings of this study are available from
the HEASARC web site: 
``https://heasarc.gsfc.nasa.gov/docs/archive.html''.

\medskip 

{\bf Code availability:} We will make all scripts used to run the MCMC analysis in Section~\ref{sec:mcmcfits} available in a google Drive folder.  The scripts use several routines in PINTofALE \url{https://hea-www.harvard.edu/PINTofALE/}.   The hardness ratio code BEHR, used in Section~\ref{sec:transitfeatures}, is available at \url{https://hea-www.harvard.edu/AstroStat/BEHR/}.



\

\bibliographystyle{mnras}
\bibliography{scibib}
\section*{Acknowledgments}
  RD would like to thank Daniel D'Orazio and Tenley Hutchinson-Smith for relevant discussions and input on the analysis of X-ray light curves. 
She would also like to thank Adam Burrows and Deepto Charkrabarty for
discussions about the models, and David Latham for a discussion about
the light curve interpretation. 
JB would like to thank Dennis Alp for discussion and help with software in the process of analyzing images and Center for Excellence in Education's Research Science Institute for enabling collaboration with RD.
TC would like to thank David Latham and his research group for discussions and guidance on transiting brown dwarfs.
VLK thanks Terry Gaetz and Piyush Sharda for useful discussions on the maths of overlapping circles.

\subsubsection*{Funding}
VLK  acknowledges support from NASA contract NAS8-03060 to the Chandra X-ray Center.
TC would like to thank 
National Science Foundation Graduate Research Fellowships Program for providing funding for this work.
NI thanks the John Harvard Distinguished Science Fellows Program for research support.

\appendix

\section{The X-ray Observations}
\subsubsection{\sl Chandra}

Between 2000 and 2018, M\,51 was observed with {\it Chandra}'s Advanced CCD Imaging Spectrometer (ACIS) a total of 16 times. Two of these observations were too short ($<2$ ks) for meaningful timing analysis and thus were ignored. We used the remaining 14 observations, which are summarized in Table \ref{tab:1}. Data were downloaded from the public archive\footnote{http://cxc.cfa.harvard.edu/} and reprocessed using standard tasks in the Chandra Interactive Analysis of Observations ({\sc ciao}) software package, version 4.11 \citep{2006SPIE.6270E..1VF}. For each observation, we found the count rate of M\,51-ULS-1. To take into account the declining sensitivity of the ACIS-S detector, particularly in the soft band, all count rates were converted to their Cycle 12 equivalent using the online tool {\sc pimms}\footnote{https://cxc.harvard.edu/toolkit/pimms.jsp} version 4.9. The corrected count rates are displayed in Table \ref{tab:2}. We extracted light curves using circular regions of $\approx4$ arcseconds, centered on M\,51 ULS, with nearby background regions at least 3 times as large. The {\it dmextract} tool was used to create background-subtracted light curves and analysis was performed using the FTOOLS task \citet{1995ASPC...77..367B} {\it lcurve}.

We utilized the spectral fitting performed by \citet{2016MNRAS.456.1859U} and reported in their Table 2 for constraints on the size of the X-ray emitting region. For full details of the spectral analysis, see \citep{}{2016MNRAS.456.1859U}.

\subsection{\sl XMM-Newton}

Of the 13 publicly-available {\it XMM-Newton} observations of M\,51, data were not taken during three observations because strong background flaring occurred. We downloaded the ten remaining observations (Table \ref{tab:1}) from NASA's High Energy Astrophysics Science Archive Research Center (HEASARC)\footnote{https://heasarc.gsfc.nasa.gov/docs/archive.html}. We reprocessed the European Photon Imaging Camera (EPIC) observations using standard tasks in the Science Analysis System (SAS) version 18.0.0 software package. Intervals of high particle background exposure were filtered out. Standard flagging routines \verb|#XMMEA_EP| and \verb|#XMMEA_EM| (along with \verb|FLAG=0| for pn) and patterns 0--4 and 0--12 were selected for pn and MOS, respectively. As with the {\it Chandra} data, we extracted the count rates of M\,51-ULS-1 for each observation, before converting them to their {\it Chandra} Cycle 12 equivalent using the {\sc pimms} tool. These corrected count rates are displayed in Table \ref{tab:2}. Light curves were extracted from circular regions with radii of 20 arcseconds, with local background regions selected to be at least three times larger.  We used the {\small SAS} tasks {\it evselect} and {\it epiclccorr} to create background-subtracted EPIC-combined light curves.

\begin{table*}
  \centering
  \begin{tabular}{lcccc}
    \hline \\
    ObsID & Observatory & Exp time & Date & Date in MJD \\
     & & (ks) &  &  \\
    (1) & (2) & (3) & (4) & (5) \\
    \hline \hline \\
    354 & Chandra & 14.86 & 2000-06-20 & 51715.34 - 51715.51\\
    1622 & Chandra & 26.81 & 2001-06-23 & 52083.78 - 52084.09\\
    112840201 & XMM & 20.916 & 2003-01-15 &52654.55	-	52654.79\\
    3932 & Chandra & 47.970 & 2003-08-07 &52858.60	-	52859.16\\
    212480801 & XMM & 49.214 & 2005-07-01 &53552.28	-	53552.85\\
    303420101 & XMM & 54.114 & 2006-05-20 &53875.27	-	53875.90\\
    303420201 & XMM & 36.809 & 2006-05-24 &53879.47	-	53879.89\\
    677980701 & XMM & 13.319 & 2011-06-07 &55719.21	-	55719.36\\
    677980801 & XMM & 13.317 & 2011-06-11 &55723.20	-	55723.35\\
    12562 & Chandra & 9.63 & 2011-06-12 &55724.29	-	55724.40\\
    12668 & Chandra & 9.99 & 2011-07-03 &55745.44	-	55745.55\\
    13813 & Chandra & 179.2 & 2012-09-09 &56179.74	-	56181.82\\
    13812 & Chandra & 157.46 & 2012-09-12 &56182.77	-	56184.59\\
    15496 & Chandra & 40.97 & 2012-09-19 &56189.39	-	56189.86\\
    13814 & Chandra & 189.85 & 2012-09-20 &56190.31	-	56192.50\\
    13815 & Chandra & 67.18 & 2012-09-23 &56193.34	-	56194.12\\
    13816 & Chandra & 73.1 & 2012-09-26 &56196.22	-	56197.06\\
    15553 & Chandra & 37.57 & 2012-10-10 &56210.03	-	56210.47\\
    19522 & Chandra & 37.76 & 2017-03-17 &57829.03	-	57829.47\\
    824450901 & XMM & 78.0 & 2018-05-13 &58251.89	-	58252.79\\
    830191401 & XMM & 98.0 & 2018-05-25 &58263.85	-	58264.99\\
    830191501 & XMM & 63.0 & 2018-06-13 &58282.07	-	58282.80\\
    830191601 & XMM & 63.0 & 2018-06-15 &58284.06	-	58284.79\\
    20988 & Chandra & 19.82 & 2018-08-31 & 58185.06	-	58553.76
  \end{tabular}
  \caption{(1) Observation ID; (2) observatory; (3) source exposure time; (4) observation date; (5) start and end of observation in modified Julian days.}
  \label{tab:1}
\end{table*}

\begin{table*}
  \centering
  \begin{tabular}{lccc}
    \hline \\
    ObsID & Average count rate & In eclipse & Out of eclipse\\
     & ($10^{-3}$ ct s$^{-1}$) & (ks) & (ks) \\
    (1) & (6) & (7) & (8) \\
    \hline \hline \\
    354 & $6.4\pm0.4$ & 0 & 14.86 \\
    1622 & $4.9\pm0.3$ & 0 & 26.81 \\
    112840201 &  $5.1\pm0.2$ & 0 & 20.916 \\
    3932 &$7.2\pm0.2$ & 0 & 47.970 \\  
    212480801& $9.5\pm0.3$ & 0 & 49.214 \\
    303420101 &$<0.6$ & \textemdash & \textemdash \\
    303420201 & $6.0\pm0.2$ & 0 & 36.809 \\
    677980701 & $2.0\pm0.3$ & 0 & 13.319 \\
    677980801 & $<3.5$ & 0 & 13.317 \\
    12562 & $<1.2$ & \textemdash & \textemdash \\
    12668 & $<1.2$ & \textemdash & \textemdash \\
    13813 & $6.1\pm0.2$ & 0 & 179.2 \\
    13812 & $7.8\pm0.2$ & 0 & 157.46 \\
    15496 & $7.6\pm0.5$ & 0 & 40.97 \\
    13814 & $12.4\pm0.3$ & $\sim$10.0 & $\sim$179.85 \\
    13815 & $13.0\pm0.5$ & $\sim$12.0 & $\sim$55.18 \\
    13816 & $0.7\pm0.2$ & 0 & $73.1$ \\
    15553 & $<0.7$ & \textemdash & \textemdash \\
    19522 & $1.1\pm0.2$ & \textemdash & \textemdash \\
    824450901 & $7.5\pm0.3$ & $\sim$44.4 & $\sim$33.6 \\
    830191401 & $<1.0$ & 0 & 98.0 \\
    830191501 & $6.0\pm0.1$ & 0 & 63.0 \\
    830191601 & $7.8\pm0.2$ & 0 & 63.0 \\
    20988 & $5.0\pm2.0$ & 0 & 19.82  
  \end{tabular}
  \caption{(1) Observation ID; (6) photon counts during active exposure; (7) time spent in eclipse; (8) time spent out of eclipse. Dashes indicate that the count rate of the source was too low for a signal to be detected.}
  \label{tab:2}
\end{table*}

\newpage
\section{Gravitational Lensing: M51-ULS-1b is not a WD}

White dwarfs can be eliminated as possible transiters of M51-ULS-1 because of the binary's youth.
Here we show that there is another reason, based on
the physics of gravitational lensing, to rule
out the possibility that M51-ULS-1b is a WD.
In the range of derived orbits ($\sim $ tens of AU; \S 1.5.1),
M51-ULS-1b  would act as a gravitational lens, increasing the amount
of light we receive from the XRS, not decreasing it. The gravitational
influence of a mass deflects light passing near it. When a mass is
dense enough to fit within a radius known as its Einstein radius, its
effect on the light reaching us from a distant point source is to {\sl
increase} the amount of light we see, rather than to cause a dimming
\citet{Einstein}. The value of an object's Einstein radius depends on
its mass, and on our distance to the lens and source. In the case we
consider, the
lens (i.e., the WD) and light source (i.e., the XRS) are
separated by a much smaller distance than the distance to the
observer. Only the distance, $a$  between the two of them plays a
significant role. The expression for the Einstein radius is
\begin{equation}
    R_E=1.16\times 10^{10} {\rm cm}\,
\Bigg(\frac{M}{M_\odot}\Bigg)^{\frac{1}{2}}
    \Bigg(\frac{a}{15\, {\rm AU}}\Bigg)^{\frac{1}{2}}
\end{equation}
Thus, any WD in the orbital range derived for M51-ULS-1b would fit
inside its Einstein ring. It would therefore serve as a
gravitational lens, increasing the numbers of X-rays we would detect
during its passage across the XRS.  It is therefore not possible that
the decrease in flux observed during the transit event was caused by
the passage of a WD.

 \section{Probability of Detecting a Transit or a Transition to Eclipse}
 
 The probability of detection, ${\mathbb P}$, is the product of a temporal factor ${\cal P}$ and a spatial factor ${\cal F}$.  Below we calculate each factor, considering in tandem both transitions to eclipse and transits.  
 
 \subsection{The Temporal Factor}
 
 Both planetary transits and transitions to stellar eclipse (an ingress or an egress) are short-lived events whose durations are many times smaller than that of typical exposures of M51.  Let $T_{obs}$ be the total time duration of exposures, and $P_{orb}$ be the orbital period. If $T_{obs}<P_{orb}$, then 
 \begin{equation}
 {\cal P} =min
 \Bigg[{\frac{T_{obs}}{P_{orb}},1}\Bigg]
 \end{equation}
If, on the other hand, $T_{obs}$ is longer than $N$ orbital periods, then ${\cal P}=1$ and, on average, $N$ ingresses and egresses will be observed.
 
 Observations of M51-ULS-1 lasted for a total of $\sim 1$~Ms$ \approx 11.6$~d.  The orbital period can be expressed as follows.
 \begin{equation}
            P_{orb}=9.6~d\, 
            \Big[\frac{a_{orb}}{50\, R_\odot}\Big]^\frac{3}{2}
            \Big[\frac{20\, M_\odot}{M_{tot}}\Big]^\frac{1}{2}
            =68~{\mathrm yr}\Big[\frac{a_{orb}}{45\, {\mathrm AU}}\Big]^\frac{3}{2}
            \Big[\frac{20\, M_\odot}{M_{tot}}\Big]^\frac{1}{2} 
 \end{equation}
 Thus, the temporal factor determining the probability of detecting an ingress
 or an egress is
 \begin{equation}
 {\cal P}_{in-eg}=min\Bigg[1.2 \, 
 \Big[\frac{T_{obs}}{1~{\mathrm Ms}}\Big]
 \Big[\frac{50\, R_\odot}{a_{orb}}\Big]^\frac{3}{2}
            \Big[\frac{M_{tot}}{20\, M_\odot}\Big]^\frac{1}{2}, 1\Bigg]
 \end{equation}
 Note that the factor on the left is larger than $1$. If, therefore, the total mass were to be $20\, M_\odot$
or larger, while at the same time, the orbital radius of the binary were to be $50\, R_\odot$ or smaller, the temporal factor would be unity for both an ingress and an egress\footnote{Note that, even if the computed value of ${\cal P}$ is larger than unity, the true value has an upper bound of unity.  A computed value  significantly larger simply signals the likelihood of detecting multiple egresses and ingresses.}.  We have not, however, determined the exact range of values of $M_{tot}$ and $a_{orb}$, so we cannot say that ${\cal P}=1$, although it does seem likely to be near unity.  The probability of detecting an ingress or an egress, however, depends as well on the orientation of the binary relative to our line of sight, which we compute in C2.
 
 In contrast, the temporal factor for the detection of a transit is small.
 \begin{equation}
 {\cal P}_{transit}=4.7\times 10^{-4} \, 
 \Big[\frac{T_{obs}}{1~{\mathrm Ms}}\Big]
 \Big[\frac{45\, {\mathrm AU}}{a_{pl}}\Big]^\frac{3}{2}
            \Big[\frac{M_{tot}}{20\, M_\odot}\Big]^\frac{1}{2} 
\end{equation}

\subsection{The Spatial Factor}

An eclipse or transit can only be detected if the orbital plane is aligned with our line of sight. The probability is
\begin{equation}
{\cal F}=\frac{(R_X+R_{ec})}{a_{orb}},
\end{equation}
where $R_X$ is the radius of the XRS and $R_{ec}$ is the radius of the eclipser. 
In the case of a stellar eclipse, $R_X<<R_{ec}$. An ingress (or egress) exhibits a sharp fall (or rise).  

The mass accretion rate in M51-ULS-1 is high (greater than roughly $10^{-6} M_\odot$~yr$^{-1}$). The donor must therefore either fill or (perhaps more likely) nearly fill its Roche lobe: $R_L < (2-3)\times R_{ec}$.
In addition, the orbital separation is likely to be $(2-3)\times R_L$.\footnote{
When the eclipsing star is fills its Roche lobe, then $a_{orb}=f(q)\, R_{ec}$, where $q=M_d/M_a$, and $f(q)$ 
is given by \citep{egg83}:
\begin{equation}
    f(q)=\frac{0.49\, q^\frac{2}{3}}{0.6\,q^\frac{2}{3} + ln\Big(1 + q^\frac{1}{3}\Big)},
\end{equation}}
We therefore write $a_{orb}=\alpha \times 6\, R_{ec},$ where the value of $\alpha$ is of order unity, and find 
\begin{equation}
    {\cal F}_{in-eg}=\frac{1}{6\, \alpha}
\end{equation}

  The probability of detecting an eclipse ingress or an eclipse egress is therefore
\begin{equation}
   {\mathbb P}_{in-eg}= {\cal F}_{in-eg} \times {\cal P}_{in-eg} = \frac{0.2}{\alpha}\, 
   \Big[\frac{T_{obs}}{1~{\mathrm Ms}}\Big]\, 
   \Big[\frac{50\, R_\odot}{a_{orb}}\Big]^\frac{3}{2}
\Big[\frac{M_{tot}}{20\, M_\odot}\Big]^\frac{1}{2}
\end{equation}

Thus, if we monitor the light curves of binaries with properties similar to those of  M51-ULS-1, there is a chance of roughly $10\%$ or $20\%$ that we will detect an ingress, and the same probability for an egress. This relatively high probability supports the hypothesis that the candidates for ingress and egress we considered in \S 4.3 are what they appear to be, and that we are therefore viewing M51-ULS-1 along its orbital plane.   This is also supported by the presence of the dip event in Figure~4 which is likely to be produced by a clump of matter associated with mass transfer passing in front of the XRS.

Viewing the system along the binary orbital plane may indicate that we
are simultaneously viewing along the plane of the planetary orbit.
Alignment is generally expected for the circumbinary disks in which planets form, and this is consistent with observations of circumbinary planets \citep{2014MNRAS.445.1731F}. A recent example of a coaligned system of circumbinary planets is Kepler-47 \citep{2019AJ.157.174O}.  
We don't know if the same should be true for planets orbiting XRBs, where formation is followed by further epochs of evolution. While some evolutionary effects, such as supernovae, may tend to disrupt alignments, others, such as mass loss in the binary plane, could enhance it.

If coplanarity  holds exactly, then when we know that the XRB is eclipsing, we are guaranteed that the spatial factor for the detection of the planetary transit is unity: 
    ${\cal F}_{trans}=1$, and
${\mathbb P}_{trans}={\cal P}_{trans}$.

However, when we study a large group of XRBs, as we have done in our archival survey, the group inevitably includes many non-eclipsing XRBs.
The probability of detecting a planetary transit is proportional to the product of the
probability that the orientation of the XRB is favorable, times the temporal factor appropriate for a candidate planet: $g\, {\mathbb P}_{transit}={\cal F}_{in-eg}\times {\cal P}_{transit}$
\begin{equation}
 {\mathbb P}_{transit}=7.8\times 10^{-5} \, g\, 
 \frac{1}{\alpha}
 \Big[\frac{T_{obs}}{1~{\mathrm Ms}}\Big]
 \Big[\frac{45\, {\mathrm AU}}{a_{pl}}\Big]^\frac{3}{2}
            \Big[\frac{M_{tot}}{20\, M_\odot}\Big]^\frac{1}{2} 
\end{equation}
Note that we have included a factor $g$ to account for the possible misalignment of the binary and planetary orbit.  

The opposite extreme corresponds to the case in which there is no correlation between the stellar and planetary orbital alignments. In such cases, 
\begin{equation}
{\cal F}_{trans}=1.2 \times 10^{-5}\,  \eta\,  \Big[\frac{(R_{X}+R_{ec})}{2\, R_J}\Big]
\Big[\frac{45\, {\rm AU}}{a_{pl}}\Big]\,
\end{equation}
Thus, if there there is no alignment, the probability of detection could be ${\cal O}(10^-5)$ smaller than if there is full alignment. The detection of a single transit would then correspond to a very large population of planet candidates.  Because the transit by M51-ULS-1b appears in a light curve that also shows evidence of transitions to and from eclipse and the possible passage of an accretion feature in front of the XRS, complete independence of the planetary and binary orbital orientations seems unlikely. We will use equation (C9), while allowing the value of $g$ to vary.

\end{document}